\documentclass[journal]{IEEEtran} 

\usepackage{graphicx,amsmath,amssymb,cite,citesort,subfigure,bm,color,algorithm} 

\usepackage{algpseudocode}

\usepackage{multicol,multirow}
\usepackage{url}
\usepackage[T1]{fontenc}

\begin{document}

\title{
  Scalable and Near-Optimal
  Discrete Phase Shift Optimization
  for
  Reconfigurable Intelligent Surfaces
  with Over 20,000 Elements
}

\author{Yuto~Hama,~\IEEEmembership{Member,~IEEE,}
  Daisuke Kitayama, 
  Kensuke Inaba, 
  Toshimori Honjo, 
  Hiroki Takesue,~\IEEEmembership{Member, IEEE},
  \\
  Naoki Ishikawa,~\IEEEmembership{Senior Member, IEEE},
  and
  Hiroyuki Takahashi,~\IEEEmembership{Member, IEEE}
  \\
  \thanks{
    Yuto Hama and Naoki Ishikawa are with the Department of Electrical and Computer Engineering,
    Yokohama National University, Yokohama 2410851, Japan
    (email: yuto.hama@ieee.org; ishikawa-naoki-fr@ynu.ac.jp).
    Daisuke Kitayama and Hiroyuki Takahashi are with the Device Technology Laboratories,
    NTT, Inc., Atsugi 2430198, Japan
    (email: daisuke.kitayama@ntt.com; hryk.takahashi@ntt.com).
    Kensuke Inaba, Toshimori Honjo, and Hiroki Takesue are with the Basic Research Laboratories,
    NTT, Inc., Atsugi 2430198, Japan
    (email: hiroki.takesue@ntt.com; toshimori.honjo@ntt.com; kensuke.inaba@ntt.com).
    {\it (Corresponding author: Yuto Hama)}
  }
}

\maketitle



\begin{abstract}
  This paper proposes a novel optimization framework for discrete phase shifts of a reconfigurable intelligent surface (RIS) using a coherent Ising machine (CIM).
  Unlike conventional methods based on iterative convex approximation or combinatorial search with exponentially increasing complexity, the CIM physically explores the solution space of Ising Hamiltonians through collective mode competition in a network of optical oscillators, enabling efficient large-scale discrete optimization.
  We formulate the RIS discrete phase optimization problem as a quadratic Ising model, which supports both binary and quaternary phase shifts by appropriately mapping quantized phase states to spin variables.
  Using a real hardware CIM, we experimentally solve quadratic optimization problems for RISs with up to 22,201 elements. The results demonstrate that the proposed method achieves physically consistent beam patterns under both line-of-sight and non-line-of-sight environments and attains the theoretical gain when transitioning from binary to quaternary phase shift.
  To further enhance scalability,
  we introduce a spin-size reduction approach that removes spins
  deterministically fixed by dominant channel components.
  This technique efficiently reduces the problem size for CIM in line-of-sight conditions
  without performance loss. 
  These results confirm that CIM-based optimization offers a practical and highly scalable solution for large RIS deployments with discrete phase shift constraints.
\end{abstract}


\begin{keywords}
  Coherent Ising machine~(CIM),
  discrete phase shift,
  Ising model,
  quadratic unconstrained binary optimization~(QUBO),
  reconfigurable intelligent surface~(RIS).
\end{keywords}

\IEEEpeerreviewmaketitle

\section{Introduction}
\label{sec:introduction}

Reconfigurable intelligent surface (RIS) has recently emerged as a transformative technology that enables the reconfiguration of wireless propagation environments by controlling a large number of passive reflecting elements. By appropriately adjusting the reflection phases, RIS can significantly enhance signal power, improve coverage, and boost both spectral and energy efficiency, making it a promising candidate for beyond 5G and 6G networks. A unique feature of RIS is that its beamforming gain increases quadratically with the number of reflecting elements~\cite{Wu2019,DiRenzo2020,Huang2019}, motivating the deployment of large RISs with tens of thousands of elements to fully exploit its potential.

While the theoretical performance of RIS is often analyzed under the assumption of \emph{continuous phase shift}, such idealization is impractical because hardware implementations typically allow only a limited set of \emph{discrete phase levels}~\cite{Wu2019,cui2014coding,kitayama2026transmissive}. Despite prior studies showing that even four quantization levels can nearly achieve the performance with continuous phase shifts~\cite{Sang2023}, the optimization of discrete phase shifts is inherently challenging. Unlike the continuous case, where convex optimization and closed-form designs are possible~\cite{Zhang2020Capacity},
the optimization of discrete phase shifts is an NP-hard problem
and the globally optimal solution is attainable only through exhaustive search~\cite{You2020},
whose complexity scales exponentially with the number of elements.

A significant body of work has explored RIS optimization under the discrete phase constraint.
In the literature,
early studies on RIS-aided communications have established fundamental design methodologies from multiple perspectives. In particular, channel estimation and passive beamforming under discrete phase shifts were investigated in~\cite{You2020}, where a progressive refinement framework was developed for efficient acquisition of cascaded channels. From a system-level viewpoint, the joint optimization of active beamforming at the access point and passive phase shifts at the RIS was studied in~\cite{guo2020weighted} via weighted sum-rate maximization for multiuser systems. Furthermore, the fundamental performance limits of RIS-aided multiple-input multiple-output~(MIMO) communications were characterized in~\cite{Zhang2020Capacity}, where the channel capacity was analyzed through joint optimization of transmit covariance and RIS reflection coefficients.

Despite these advances, most existing works rely on idealized hardware assumptions in which each RIS element is modeled with unit-modulus reflection independent of its phase shift. However, practical RIS implementations exhibit non-negligible hardware impairments, where the reflection amplitude depends on the phase shift due to circuit-level losses. To address this issue, a practical phase shift model incorporating amplitude–phase coupling was proposed in~\cite{Abeywickrama2020}, demonstrating that ignoring such hardware effects may lead to noticeable performance degradation.
Another fundamental challenge in RIS design arises from the discrete nature of phase configurations. The resulting optimization problem is inherently non-convex and combinatorial. To tackle this, \cite{xiong2024optimal} reformulated the discrete beamforming problem as an inner product maximization problem and proposed an efficient divide-and-sort algorithm that achieves global optimality with significantly reduced complexity compared to exhaustive search. More recently, \cite{lu2024optimal} extended this line of work to more realistic scenarios with non-uniform phase quantization across RIS elements and developed a partition-and-traversal algorithm to obtain globally optimal solutions under heterogeneous discrete constraints.


However,
existing methods inherently suffer from fundamental limitations in terms of scalability and practical applicability. Most approaches rely on iterative optimization procedures whose convergence behavior and computational cost become increasingly prohibitive as the problem size grows. At the same time, methods that directly address discrete phase constraints typically involve combinatorial search over exponentially large solution spaces, which quickly becomes intractable for large-scale RISs. Moreover, many existing frameworks are not fully compatible with hardware-constrained discrete phase models when the number of reflecting elements is large, further limiting their applicability in practical deployments.
These challenges highlight the need for fundamentally different optimization paradigms that can efficiently handle large-scale discrete optimization problems beyond the scope of conventional iterative or combinatorial methods.

In this paper, we introduce a novel approach for discrete phase shift optimization based on the \emph{coherent Ising machine (CIM)}. CIMs are unconventional computing architectures that solve large-scale combinatorial optimization problems by exploiting the collective dynamics of coupled oscillators~\cite{Inagaki2016,McMahon2016,Yamamoto2017,Cen2022}.
Notably, CIMs inherently support all-to-all connectivity among spins,
which provides a significant advantage over quantum annealers~\cite{hamerly2019experimental}.
By mapping the RIS discrete phase shift problem into an Ising Hamiltonian,
we leverage the collective mode competition inherent in CIM to efficiently obtain near-optimal solutions~\cite{cummins2025ising},
even when the number of RIS elements reaches tens of thousands. Unlike conventional digital algorithms, the CIM-based approach enables scalable optimization under strict discrete constraints, offering a promising solution to the fundamental bottleneck of RIS implementation.

The main contributions of this work are as follows:
\begin{itemize}
\item We propose a novel optimization framework for discrete phase shifts by formulating the RIS-assisted beamforming problem as an Ising Hamiltonian and solving it with a coherent Ising machine~(CIM), enabling highly efficient physical parallel search for large-scale combinatorial optimization.
\item We show that the proposed CIM-based approach supports both binary and quaternary discrete phase shifts and demonstrate that the CIM achieves physically consistent beam patterns and the theoretical gain improvement when transitioning from binary to quaternary phase shift.
\item We introduce a spin-size reduction technique that exploits the dominance of the line-of-sight~(LoS) terms to deterministically remove trivial spins. The simulation result demonstrates that this technique reduces the number of spins without any performance loss.
\item We conduct performance evaluations using a real hardware CIM and validate that the proposed approach successfully optimizes large-scale RISs with 22,201 elements.
It is important to note that the number of spins required by the CIM scales linearly with the number of RIS elements, indicating that the proposed method is scalable to large-scale RISs.
To the best of our knowledge, this is the first work to introduce the CIM for discrete phase shift design for RISs.
\end{itemize}


The remainder of this paper is organized as follows.
Section~II introduces the system model.
Section~III presents the proposed CIM-based optimization method together with the formulation of the discrete phase optimization problem.
Section~IV proposes the spin size reduction approach for LoS channels.
The simulation results are given in Section~V,
where the achievable performance of our proposed method is evaluated in terms of the channel gain
by using a real hardware CIM.
Section~VI concludes this work.

\section{System Model}
\label{sec:system}

\begin{figure}[tb]
  \centering
  \includegraphics[width = 0.8\hsize, clip]{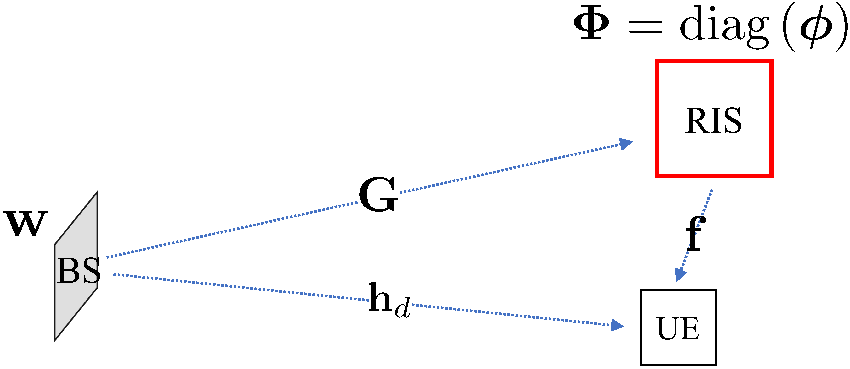}
  \caption{
    The system model of RIS-assisted single user downlink communication.
  }
  \label{system_model}
\end{figure}

In this section,
we describe the system model of RIS-assisted cellular networks
shown in Fig.~\ref{system_model}.
Throughout this work,
we consider a downlink single-user multiple-input single-output (SU-MISO) system,
where a base station (BS) with~$N_\text{BS}$ antennas communicates with a single-antenna user equipment~(UE)
with the assistance of a single RIS with~$N_\text{RIS}$ elements.

\subsection{Symbol Representation}

Let $s \in \mathbb{C}$ and $r \in \mathbb{C}$ denote the transmit and receive symbols, respectively,
where we assume that the transmit symbol is normalized to unit power,
i.e.,~$E \left\{ \left| s \right|^2 \right\} = 1$.
The channel between BS and UE,
denoted by~$\mathbf{h} \in \mathbb{C}^{N_\text{BS} \times 1}$,
is composed of two different channels of
direct link between BS and UE
denoted by~$\mathbf{h}_{d} = \left( h_{d,1}, \cdots, h_{d,N_\text{BS}} \right)^{\top} \in \mathbb{C}^{N_\text{BS} \times 1}$
and reflecting link via RIS
denoted by~$\mathbf{h}_{r} = \left( h_{r,1}, \cdots, h_{r,N_\text{BS}} \right)^{\top} \in \mathbb{C}^{N_\text{BS} \times 1}$.
Let $\boldsymbol\Phi = \text{diag} \left( \boldsymbol\phi \right) = \text{diag} \left( \phi_{1}, \cdots, \phi_{N_\text{RIS}} \right) \in \mathbb{C}^{N_\text{RIS} \times N_\text{RIS}}$ denote
the diagonal reflection coefficient matrix of the RIS,
and its $n$th element is given by $\phi_{n} \triangleq \beta_{n} e^{j\theta_{n}}$,
where $\beta_{n} \in [0,1]$ and $\theta_{n} \in [0, 2\pi)$ denote
the reflection amplitude and phase shift
performed at the $n$th element of the RIS,
respectively.
The channel of reflecting link is expressed as
\begin{align}
  \mathbf{h}_{r}^{\top}
  &=
  \mathbf{f}^{\top} \boldsymbol\Phi \mathbf{G},
  \label{eq:h_r}
\end{align}
where $\mathbf{f} = \left( f_{1}, \cdots, f_{N_\text{RIS}} \right)^{\top} \in \mathbb{C}^{N_\text{RIS} \times 1}$ is the channel from the RIS to the UE,
and $\mathbf{G} \in \mathbb{C}^{N_\text{RIS} \times N_\text{BS}}$ is the channel matrix from the BS to the RIS.
Hence,
the channel from the BS to the UE is given by
\begin{align}
  \mathbf{h}^{\top}
  &=
  \mathbf{h}_{d}^{\top} + \mathbf{h}_{r}^{\top}
  =
  \mathbf{h}_{d}^{\top} +
  \mathbf{f}^{\top} \boldsymbol\Phi \mathbf{G}
  .
  \label{eq:h}
\end{align}
Therefore,
the received symbol at the UE is expressed as
\begin{align}
  r &=
  \mathbf{h}^{\top} \mathbf{w} s + z
  =
  \left( \mathbf{h}_{d}^{\top} + \mathbf{h}_{r}^{\top} \right) \mathbf{w} s + z
  \nonumber\\&
  =
  \left( \mathbf{h}_{d}^{\top} + \mathbf{f}^{\top} \boldsymbol\Phi \mathbf{G} \right) \mathbf{w} s
  + z
  ,
  \label{eq:r_k}
\end{align}
where $\mathbf{w} = \left( w_1, \cdots, w_{N_\text{BS}} \right)^{\top} \in \mathbb{C}^{N_\text{BS} \times 1}$ is the beamforming~(BF) weight vector performed at the BS,
and $z$ represents an additive white Gaussian noise~(AWGN) term,
which follows complex Gaussian distribution with zero mean and complex variance of~$N_0$,
i.e.,~$z \sim \mathcal{CN} \left( 0, N_0 \right)$.

\subsection{Channel Model}

In this work, we consider a free-space propagation environment where the BS, the UE,
and the RIS are located in a free space without any additional scatterers.
Under this assumption,
all channel components are modeled solely based on geometric distances and free-space propagation loss.

\subsubsection{Direct Channel}

We first consider the direct path between the BS and the UE.
Using the free-space path-loss model,
the channel from the $k$th BS antenna to the UE is expressed as
\begin{align}
h_{{d},k}
= \frac{\lambda}{4\pi d_{\mathrm{BS\text{-}UE},k}}
\exp\!\left(-j \frac{2\pi}{\lambda} d_{\mathrm{BS\text{-}UE},k}\right),
\label{eq:hd_element}
\end{align}
where $d_{\mathrm{BS\text{-}UE},k}$ is the distance between the $k$th BS antenna and the UE, and $\lambda$ is the wavelength.

\subsubsection{RIS-Assisted Reflected Channel}

For the reflected path via the RIS,
we adopt a physically consistent RIS channel model
that is aligned with experimentally validated formulations presented in~\cite{kitayama2023alignment}.
The overall reflected channel is determined by the BS-RIS channel matrix $\mathbf{G} \in \mathbb{C}^{N_\text{RIS} \times N_\text{BS}}$ and the RIS-UE channel vector $\mathbf{f} \in \mathbb{C}^{N_\text{RIS} \times 1}$.

Let $G_{n,k}$ denote the channel between the $k$th BS antenna and the $n$th RIS element,
which corresponds to the $(n,k)$th element of~$\mathbf{G}$.
According to the Friis transmission formula, the element-wise channel is
\begin{align}
G_{n,k}
= \sqrt{\frac{A_{\mathrm{BS},n,k}}{4\pi d_{\mathrm{BS\text{-}RIS},n,k}}}
\exp\!\left(-j \frac{2\pi}{\lambda} d_{\mathrm{BS\text{-}RIS},n,k}\right),
\label{eq:G_nk}
\end{align}
where $d_{\mathrm{BS\text{-}RIS},n,k}$ is the distance between the $k$th BS antenna and the $n$th RIS element,
and $A_{\mathrm{BS},n,k}$ is the effective aperture area of the $n$th RIS element as viewed from the $k$th BS antenna.

Similarly, the channel between the $n$th RIS element and the UE is given by
\begin{align}
f_{n}
= \sqrt{\frac{A_{\mathrm{UE},n}}{4\pi d_{\mathrm{RIS\text{-}UE},n}}}
\exp\!\left(-j \frac{2\pi}{\lambda} d_{\mathrm{RIS\text{-}UE},n}\right),
\label{eq:f_j}
\end{align}
where $d_{\mathrm{RIS\text{-}UE},n}$ denotes the distance between the $n$th RIS element and the UE, and $A_{\mathrm{UE},n}$ is the effective aperture area of the $n$th RIS element as observed from the UE.

Combining \eqref{eq:G_nk} and \eqref{eq:f_j},
the cascaded channel from the $k$th BS antenna to the UE via the $n$th RIS element with
reflection coefficient~$\phi_n = \beta_n e^{j\theta_n}$
is expressed as
\begin{align}
f_n \, \phi_n \, G_{n,k}
&=
\sqrt{\frac{A_{\mathrm{UE},n}}{4\pi d_{\mathrm{RIS\text{-}UE},n}}}
\exp\!\left(-j \frac{2\pi}{\lambda} d_{\mathrm{RIS\text{-}UE},n}\right)
\nonumber\\
&\,\,\,\,\,\,
\times
\phi_n
\sqrt{\frac{A_{\mathrm{BS},n,k}}{4\pi d_{\mathrm{BS\text{-}RIS},n,k}}}
\exp\!\left(-j \frac{2\pi}{\lambda} d_{\mathrm{BS\text{-}RIS},n,k}\right)
\nonumber\\
&=
\frac{\beta_n e^{j\theta_n} \sqrt{A_{\mathrm{UE},n} A_{\mathrm{BS},n,k}}}
     {4\pi \, \sqrt{d_{\mathrm{RIS\text{-}UE},n} \, d_{\mathrm{BS\text{-}RIS},n,k}}}
\nonumber\\
&\,\,\,\,\,\,
\times
\exp\!\left(
-j \frac{2\pi}{\lambda}
\big( d_{\mathrm{BS\text{-}RIS},n,k} + d_{\mathrm{RIS\text{-}UE},n} \big)
\right).
\label{eq:cascaded_channel}
\end{align}

\subsection{Discrete Phase Shift}

Throughout this work,
we consider discrete phase shifts of RIS.
Thus,
each element on the RIS keeps amplitude unchanged and controls the phase only,
i.e.,~$\beta_{n} = 1$ with $n\in\left\{ 1, \cdots, N_\text{RIS} \right\}$.
Furthermore,
the phase at each element~$\theta_n$ is chosen from
a set of finite discrete phase candidates denoted by~$\mathcal{D}_L$,
where $L$ represents the number of phase shift levels.
Thus,
$\left| \mathcal{D}_L \right| = L$ holds
with $\left| \cdot \right|$ representing the cardinality. 
The set of discrete phases is expressed as
\begin{align}
  \mathcal{D}_L = \left\{ 0, \frac{2\pi}{L}, \cdots, \left( L-1 \right) \frac{2\pi}{L} \right\}
  .
  \label{eq:D}
\end{align}
Therefore,
the discrete phase vector of RIS
is given by
\begin{align}
  \boldsymbol\theta =
  \left( \theta_1, \cdots, \theta_{N_\text{RIS}} \right)
  \in {\mathcal{D}_L}^{N_\text{RIS}}
  .
  \label{eq:discrete_phase_vector}
\end{align}
As a result,
the reflection coefficient corresponding to the $n$th element is reduced to
\begin{align}
  \phi_{n}
  &=
  e^{j\theta_{n}}
  ,
  \qquad
  \theta_{n} \in \mathcal{D}_L
  .
\end{align}

The number of combinations for discrete phase shifts of RIS with $N_\text{RIS}$ elements
is given by~$\left| {\mathcal{D}_L}^{N_\text{RIS}} \right| = L^{N_\text{RIS}}$.
Since the computational complexity required for the exhaustive search grows exponentially,
the optimization for a large~$N_\text{RIS}$ is unattainable even with the lowest case of~$L = 2$.

\begin{table}[tb] 
  \centering      
  \caption{Power loss by discrete phase shifts~\cite{wu2019beamforming}} 
  \label{discrete_loss} 
  \begin{tabular}{|c||c|c|c|c|} 
    \hline
    Phase shift level~$L$ & 2 & 4 & 8 & $\infty$ \\ \hline
    Power loss & 3.9~dB & 0.9~dB & 0.2~dB & 0~dB \\ \hline
  \end{tabular}
\end{table}

\begin{figure}[tb]
  \centering
  \includegraphics[width = \hsize, clip]{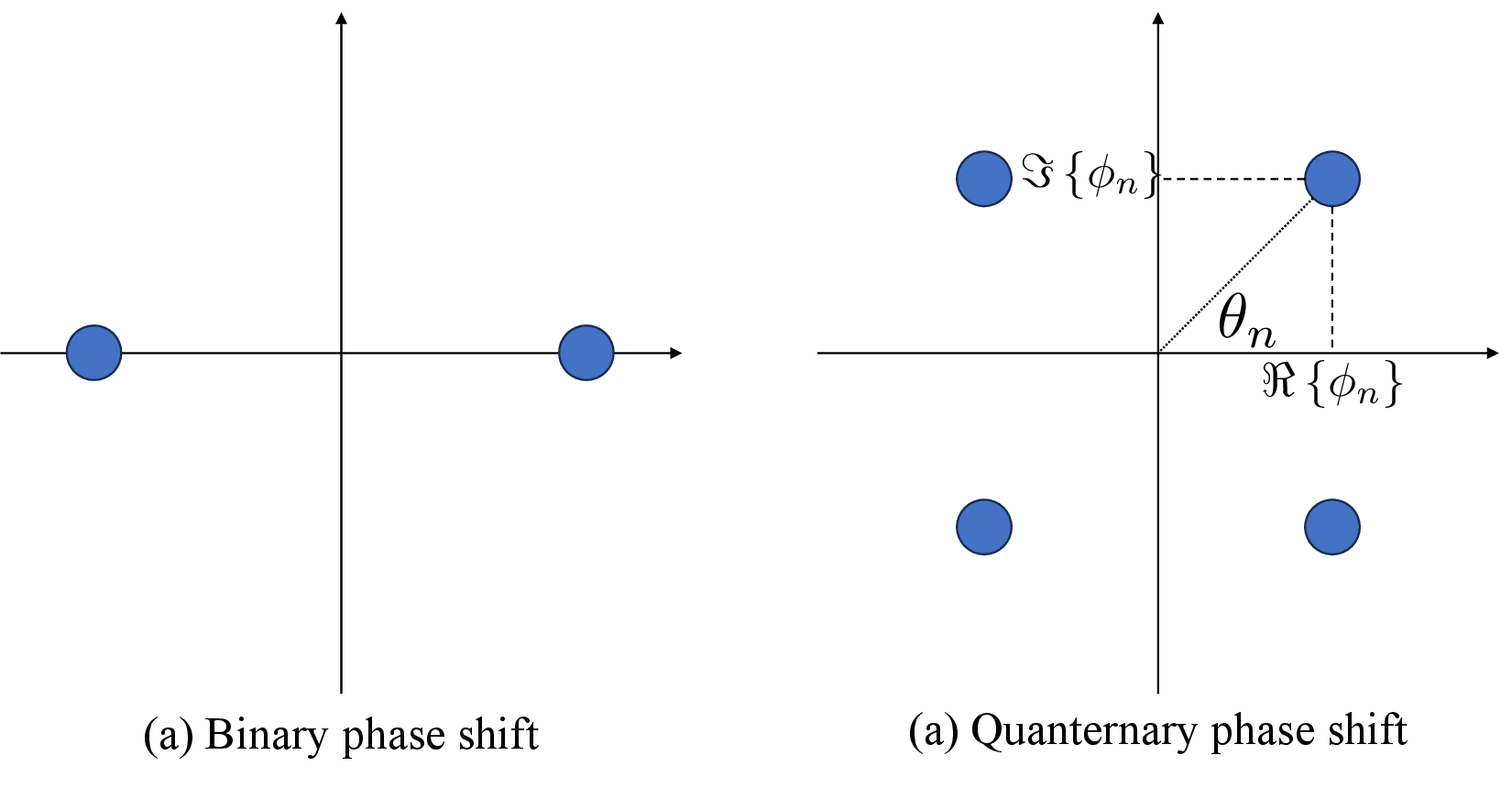}
  \caption{
    Discrete phase shifts with phase shift levels of~$L = 2$ (left) and $L=4$ (right).
  }
  \label{phase_set}
\end{figure}

Performance loss caused by discrete phase shift compared to continuous counterpart is
given in Table~\ref{discrete_loss}~\cite{wu2019beamforming}.
As given,
the power loss is sufficiently suppressed to $0.9$~dB at $L = 4$,
and it becomes nearly negligible for~$L > 4$.
Therefore,
throughout this work,
we focus only on the practical cases of~$L = 2$ and~$4$ for simplicity as follows,
and these phase sets are illustrated in Fig.~\ref{phase_set}.

\subsubsection{Binary Phase Shift ($L = 2$)}

We first consider binary phase shifts of RIS,
i.e.,~$L = 2$.
The set of discrete phase candidates is given by
\begin{align}
  \mathcal{D}_2 = \left\{ 0, \pi \right\}
  .
  \label{eq:D_2}
\end{align}
Hence,
the $n$th coefficient on the RIS~$\phi_{n}$ is chosen from the set~$\mathcal{P}_2$
(i.e.,~$\phi_{n} \in \mathcal{P}_2$),
which is expressed as
\begin{align}
  \mathcal{P}_2
  \triangleq
  \left\{ e^{j 0}, e^{j \pi} \right\}
  =
  \left\{ 1, -1 \right\}
  .
  \label{eq:P_2}
\end{align}


\subsubsection{Quaternary Phase Shift ($L = 4$)}

We now consider the quaternary case,
i.e.,~$L = 4$.
For the sake of analysis in Section~\ref{sec:cim_quaternary},
we apply the phase rotation of~$\pi/4$ to~\eqref{eq:discrete_phase_vector}.
The set of discrete phase candidates for quaternary phase shift is given by
\begin{align}
  \mathcal{D}_{4:\text{rotated}} =
  \left\{ \frac{\pi}{4}, \frac{3}{4}\pi, \frac{5}{4}\pi, \frac{7}{4}\pi \right\}.
  \label{eq:D_4}
\end{align}
As a result,
the $n$th coefficient on the RIS~$\phi_{n}$ is chosen from the set~$\mathcal{P}_{4}$,
which is expressed as
\begin{align}
  &\mathcal{P}_4
  \triangleq
  \left\{ e^{j \frac{\pi}{4}}, e^{j \frac{3\pi}{4}}, e^{j \frac{5\pi}{4}}, e^{j \frac{7\pi}{4}}  \right\}
  \nonumber\\&
  =
  \left\{ \frac{1}{\sqrt{2}} \left( 1 + j \right), \frac{1}{\sqrt{2}} \left( -1 + j \right), \frac{1}{\sqrt{2}} \left( -1 - j \right), \frac{1}{\sqrt{2}} \left( 1 - j \right) \right\}
  .
  \label{eq:P_4}
\end{align}

\section{Discrete Phase Shift Optimization Based on Coherent Ising Machine}

In this section,
we propose an optimization approach based on CIM
for discrete phase shifts of RIS.
To this end,
we first describe the principle of CIM together with the Ising model.
Then,
we derive that the discrete phase shifts of RISs
can be formulated as an Ising model that makes it compatible with the CIM.

\subsection{Principle of Coherent Ising Machine}

The CIM is a physics-inspired computing architecture that efficiently solves large-scale combinatorial optimization problems by searching for the ground state of the Ising Hamiltonian.
The Ising model represents a system of binary spins interacting with each other through
a coupling coefficient and subject to an external magnetic field.
Let $\sigma_i \in \{+1, -1\}$ denote the $i$th Ising spin.
The total energy of the system, or Ising Hamiltonian, is defined as
\begin{equation}
H_{\mathrm{Ising}}(\boldsymbol{\sigma}) =
\sum_{i=1}^{N}\sum_{j=1}^{N} J_{ij}\sigma_i\sigma_j
+ \sum_{i=1}^{N} \lambda_i\sigma_i,
\label{eq:ising_hamiltonian}
\end{equation}
where $\boldsymbol{\sigma} = [\sigma_1, \sigma_2, \ldots, \sigma_{N}]^\top \in \{ +1,-1 \}^{N\times1}$ denotes the spin configuration, $J_{ij} \in \mathbb{R}$ is the mutual interaction between the $i$th and $j$th spins, and $\lambda_i \in \mathbb{R}$ is the external magnetic field of the $i$th spin.
By introducing a vector $\boldsymbol\lambda \in \mathbb{R}^{N\times1}$, whose $i$th element is given by~$\lambda_i$
and a matrix~$\mathbf{J} \in \mathbb{R}^{N\times N}$, whose $(i,j)$th element is given by~$J_{ij}$,
the Ising Hamiltonian in~\eqref{eq:ising_hamiltonian} can be expressed in vector form as
\begin{equation}
H_{\mathrm{Ising}}(\boldsymbol{\sigma}) =
\boldsymbol\sigma^\top 
\mathbf{J}
\boldsymbol\sigma
+
\boldsymbol\lambda^\top \boldsymbol\sigma
.
\label{eq:ising_hamiltonian_vec}
\end{equation}

The objective of the CIM is to obtain the spin configuration $\hat{\boldsymbol{\sigma}}$ that minimizes $H_{\mathrm{Ising}}(\boldsymbol{\sigma})$,
and this optimization problem can be formulated as
\begin{subequations}
\begin{align}
  E
  &=
  \min_{\boldsymbol\sigma}
  H_\text{Ising} \left( \boldsymbol\sigma \right),
  \label{eq:optimization_CIM}
  \\
  &\text{s.t.} \quad
  \boldsymbol\sigma \in \left\{ +1, -1 \right\}^N
  .
  \label{eq:cim}
\end{align}
\label{eq:cim_opt}
\end{subequations}

\subsection{Problem Formulation}

In this work,
we consider the optimization problem
so as to maximize the capacity
subject to the discrete phase shift constraints in~\eqref{eq:D}.

From~\eqref{eq:r_k},
the signal-to-noise ratio~(SNR) for the given channel~$\mathbf{h}$ is defined as
\begin{align}
    \gamma \left( \mathbf{h} \right)
    &=
    \frac{\left| \mathbf{h}^\top \mathbf{w} \right|^2}{E \left\{ \left| z \right|^2 \right\}}
    =
    \frac{\left| \left( \mathbf{h}_d^\top + \mathbf{f}^\top \boldsymbol{\Phi} \mathbf{G} \right) \mathbf{w} \right|^2}{N_0}
    .
    \label{eq:snr}
\end{align}
Thus,
the capacity is expressed as
\begin{align}
    C
    &=
    \log_2 \left( 1 + \gamma \left( \mathbf{h} \right) \right)
    =
    \log_2 \left( 1 + \frac{\left| \mathbf{h}^\top \mathbf{w} \right|^2}{N_0} \right)
    \nonumber\\&
    =
    \log_2 \left( 1 + \frac{\left| \left( \mathbf{h}_d^\top + \mathbf{f}^\top \boldsymbol{\Phi} \mathbf{G} \right) \mathbf{w} \right|^2}{N_0} \right)
    .
    \label{eq:capacity}
\end{align}
Throughout this work,
we consider discrete phase shifts of the RIS in terms of capacity maximization.
Therefore,
the optimization problem can be formulated as
\begin{subequations}
\begin{align}
  E
  &\triangleq
  \max_{\mathbf{w}, \boldsymbol{\theta}} C
  \nonumber\\
  &=
  \max_{\mathbf{w}, \boldsymbol{\theta}}
  \log_2 \left( 1 +
  \frac
  {\left| \mathbf{h}^\top \mathbf{w} \right|^2}
  {N_0}
  \right)
  ,
  \label{eq:E}
  \\
  &\text{s.t.} \quad
  \left\| \mathbf{w} \right\|^2 \leq 1,
  \\
  &\text{s.t.} \quad
  \boldsymbol\theta \in {\mathcal{D}_L}^{N_\text{RIS}}
  ,
\end{align}
\label{eq:opt}
\end{subequations}
where $\left\| \mathbf{a} \right\| \triangleq \sqrt{\sum_{n=1}^{N} \left| a_n \right|^2 }$ denotes
the $\ell^2$-norm of the vector~$\mathbf{a} \in \mathbb{C}^{N\times1}$.

It is important to note that
the optimal BF weight vector at BS
is always given by the maximum ratio transmission 
independent of any phase shifts of RIS~$\boldsymbol\theta$
in single-user MISO system~\cite{telatar,wu2019beamforming}.
Thus,
the BF weight vector at the BS is expressed as
\begin{align}
  \mathbf{w}
  &=
  \frac{\left(\mathbf{h}^\top\right)^\text{H} }{\left\| \mathbf{h}^\top \right\|}
  =
  \frac{\left( \mathbf{h}_{d}^\top + \mathbf{f}^\top \boldsymbol\Phi \mathbf{G} \right)^\text{H}}{\left\| \mathbf{h}_{d}^\top + \mathbf{f}^\top \boldsymbol\Phi \mathbf{G} \right\|}
  .
  \label{eq:w}
\end{align}

Therefore,
the optimization problem in~\eqref{eq:opt} can be rewritten as
\begin{subequations}
\begin{align}
  E
  &=
  \max_{\boldsymbol{\theta}}
  \left\| \mathbf{h}^\top \right\|^2
  ,
  \label{eq:E_2}
  \\
  &\text{s.t.} \quad
  \boldsymbol\theta \in {\mathcal{D}_L}^{N_\text{RIS}}.
\end{align}
\label{eq:RIS_opt}
\end{subequations}

\subsection{Binary Phase Shift}

We first consider the optimization for binary phase shift.
In this case, from~\eqref{eq:P_2},
the $n$th reflection coefficient is given by
$\phi_n = \{ +1,-1 \}$.
Hence,
the reflection coefficient vector~$\boldsymbol\phi = \left( e^{j\theta_1}, \cdots, e^{j\theta_{N_\text{RIS}}} \right)^\top \in \{ +1,-1 \}^{N_\text{RIS} \times 1}$
directly corresponds to
the spin vector~$\boldsymbol\sigma$ in Ising model.
Therefore,
\eqref{eq:RIS_opt} can be rewritten as
\begin{subequations}
\begin{align}
  E
  &=
  \min_{\boldsymbol{\phi}}
  \left[
  -
  \left\| \mathbf{h}_{d}^\top + \mathbf{f}^\top \boldsymbol\Phi \mathbf{G} \right\|^2
  \right]
  \nonumber\\
  &=
  \min_{\boldsymbol{\phi}}
  \left[
  -
  \left\| \mathbf{h}_{d}^\top + \mathbf{f}^\top \text{diag} \left( \boldsymbol\phi \right) \mathbf{G} \right\|^2
  \right]
  ,
  \\
  &\text{s.t.} \quad
  \boldsymbol\phi \in \{+1, -1\}^{N_\text{RIS} \times 1}.
\end{align}
\label{eq:optimization_binary}
\end{subequations}
From~\eqref{eq:h_r},
the channel corresponding to reflecting link can be rewritten as
\begin{align}
  \mathbf{h}_{r}^\top
  &=
  \mathbf{f}^\top \boldsymbol\Phi \mathbf{G}
  =
  \mathbf{f}^\top \text{diag} \left( \boldsymbol\phi \right) \mathbf{G}
  \nonumber\\&
  =
  \boldsymbol\phi^\text{H} \text{diag} \left( \mathbf{f} \right) \mathbf{G}
  \triangleq
  \boldsymbol\phi^\text{H} \mathbf{V},
  \label{eq:reflect_channel}
\end{align}
where $\mathbf{V} \triangleq \text{diag} \left( \mathbf{f} \right) \mathbf{G} \in \mathbb{C}^{N_\text{RIS}\times N_\text{BS}}$.
Thus,
we have
\begin{align}
  \left\| \mathbf{h}^\top \right\|^2
  &=
  \left\| \mathbf{h}_{d}^\top + \mathbf{h}_r^\top \right\|^2
  =
  \left\| \mathbf{h}_{d}^\top + \boldsymbol\phi^\text{H} \mathbf{V} \right\|^2
  \nonumber\\&
  =
  \left( \mathbf{h}_{d}^\top + \boldsymbol\phi^\text{H} \mathbf{V} \right)
  \left( \mathbf{h}_{d}^\top + \boldsymbol\phi^\text{H} \mathbf{V} \right)^\text{H}
  \nonumber\\&
  =
  \left\| \mathbf{h}_{d}^\top \right\|^2
  +
  \boldsymbol\phi^\text{H} \mathbf{V} \mathbf{V}^\text{H} \boldsymbol\phi
  +
  2 \Re \left\{ \mathbf{h}_d^\top \mathbf{V}^\text{H} \boldsymbol\phi \right\}
  .
  \label{eq:h_norm}
\end{align}
Since $\left\| \mathbf{h}^\top \right\|^2 \in \mathbb{R}$ is a real value
and each element of $\boldsymbol\phi \in \{ +1,-1 \}^{N_\text{RIS}}$ takes real values,
\eqref{eq:h_norm} can be rewritten as
\begin{align}
  \left\| \mathbf{h}^\top \right\|^2
  &=
  \left\| \mathbf{h}_{d}^\top \right\|^2
  +
  \boldsymbol\phi^\top \Re \left\{ \mathbf{V} \mathbf{V}^\text{H} \right\} \boldsymbol\phi
  +
  2 \Re \left\{ \mathbf{h}_d^\top \mathbf{V}^\text{H} \right\} \boldsymbol\phi
  .
  \label{eq:h_norm_real}
\end{align}

Since the first term~$\left\| \mathbf{h}_{d}^\top \right\|^2$ in~\eqref{eq:h_norm_real}
is independent of the reflecting coefficient vector~$\boldsymbol\phi$,
the optimization problem for binary phase shift is expressed
from~\eqref{eq:optimization_binary} and~\eqref{eq:h_norm_real} as
\begin{subequations}
\begin{align}
  E
  &=
  \min_{\boldsymbol{\phi}}
  \left[
    \boldsymbol\phi^\top
  \left( -\Re\left\{ \mathbf{V} \mathbf{V}^\text{H}  \right\} \right)
  \boldsymbol\phi
  - 2 \Re \left\{ \mathbf{h}_d \mathbf{V}^\text{H} \right\} \boldsymbol\phi
  \right]
  \nonumber\\&
  \triangleq
  \min_{\boldsymbol{\phi}}
  \left[
    \boldsymbol\phi^\top
    \mathbf{J}_\text{b}
  \boldsymbol\phi
  +
  \boldsymbol{\lambda}^\top_\text{b}
  \boldsymbol\phi
  \right]
  ,
  \\
  &\text{s.t.} \quad
  \boldsymbol\phi \in
  \left\{ +1, -1 \right\}^{N_\text{RIS} \times 1}
  ,
  \label{eq:E_su_binary}
\end{align}
\end{subequations}
where
the mutual interaction matrix and
the external magnetic field vector
for binary phase shift on RIS,
denoted by~$\mathbf{J}_\text{b} \in \mathbb{R}^{N_\text{RIS} \times N_\text{RIS}}$ and~$\boldsymbol{\lambda}_\text{b} \in \mathbb{R}^{N_\text{RIS} \times1}$, respectively,
are expressed as
\begin{align}
  \mathbf{J}_\text{b}
  &=
  -
  \Re \left\{ \mathbf{V} \mathbf{V}^\text{H} \right\}
  \nonumber\\&
  =
  -
  \Re \left\{
  \text{diag} \left( \mathbf{f} \right) \mathbf{G} \mathbf{G}^\text{H} \text{diag} \left( \mathbf{f} \right)^\text{H}
  \right\}
  \label{eq:J_b}
  ,
  \\
  \boldsymbol{\lambda}^\top_\text{b}
  &=
  -2 \Re \left\{ \mathbf{h}_d^\top \mathbf{V}^\text{H} \right\}
  \nonumber\\&
  =
  -2 \Re \left\{ \mathbf{h}_d^\top \mathbf{G}^\text{H} \text{diag} \left( \mathbf{f} \right)^\text{H} \right\}
  .
  \label{eq:lambda_b}
\end{align}

\subsection{Quaternary Phase Shift}
\label{sec:cim_quaternary}

We next consider quaternary phase shift case.
It is important to note that
the reflection coefficient $\phi_n$ can be represented in binary
by dividing it into real and imaginary parts
even though the reflection coefficient $\phi_n$ itself is not binary.
From~\eqref{eq:P_4},
the real and imaginary parts of the reflection coefficient~$\phi_n$ are given by
\begin{align}
  \Re \left\{ \phi_n \right\} &\in \left\{ +\frac{1}{\sqrt{2}}, -\frac{1}{\sqrt{2}} \right\}
  =
  \frac{1}{\sqrt{2}} \left\{ +1, -1 \right\}
  ,
  \\
  \Im \left\{ \phi_n \right\} &\in \left\{ +\frac{1}{\sqrt{2}}, -\frac{1}{\sqrt{2}} \right\}
  =
  \frac{1}{\sqrt{2}} \left\{ +1, -1 \right\}
  .
\end{align}

Equation~\eqref{eq:h_norm} can be rewritten by separating
the reflecting coefficient vector~$\boldsymbol\phi$
into its real and imaginary parts as
\begin{align}
  \left\| \mathbf{h}^\top \right\|^2
  &=
  \left\| \mathbf{h}_{d}^\top \right\|^2
  +
  \boldsymbol\phi^\text{H} \mathbf{V} \mathbf{V}^\text{H} \boldsymbol\phi
  +
  2 \Re \left\{ \mathbf{h}_d^\top \mathbf{V}^\text{H} \boldsymbol\phi \right\}
  \nonumber\\&
  =
  \left\| \mathbf{h}_{d}^\top \right\|^2
  \nonumber\\&
  \quad
  +
  \left(
  \Re \left\{ \boldsymbol\phi^\text{H} \right\} \Re \left\{ \mathbf{V} \mathbf{V}^\text{H} \right\} \Re \left\{ \boldsymbol\phi \right\}
  \right.
  \nonumber\\&
  \quad
  -
  \Im \left\{ \boldsymbol\phi^\text{H} \right\} \Re \left\{ \mathbf{V} \mathbf{V}^\text{H} \right\} \Im \left\{ \boldsymbol\phi \right\}
  \nonumber\\&
  \quad
  -
  \Re \left\{ \boldsymbol\phi^\text{H} \right\} \Im \left\{ \mathbf{V} \mathbf{V}^\text{H} \right\} \Im \left\{ \boldsymbol\phi \right\}
  \nonumber\\&
  \quad
  \left.
  -
  \Im \left\{ \boldsymbol\phi^\text{H} \right\} \Im \left\{ \mathbf{V} \mathbf{V}^\text{H} \right\} \Re \left\{ \boldsymbol\phi \right\}
  \right)
  \nonumber\\&
  \quad
  +
  \left(
  2 \Re \left\{ \mathbf{h}_d^\top \mathbf{V}^\text{H} \right\} \Re \left\{ \boldsymbol\phi \right\}
  -
  2 \Im \left\{ \mathbf{h}_d^\top \mathbf{V}^\text{H} \right\} \Im \left\{ \boldsymbol\phi \right\}
  \right)
  .
  \label{eq:h_norm_q}
\end{align}

Let $\boldsymbol\phi \in \left\{ +1+1j, -1+1j, -1-1j, +1-1j \right\}^{N_\text{RIS} \times 1}$
be separated into its real and imaginary parts, and define a new vector
$\bar{\boldsymbol\phi} = \left( \Re\{ \boldsymbol\phi \}^\top, \Im\{ \boldsymbol\phi \}^\top \right)^\top \in \left\{ +1,-1 \right\}^{2N_\text{RIS} \times1}$
of twice length.
From~\eqref{eq:h_norm_q},
the optimization problem for quaternary phase shift can be formulated as
\begin{subequations}
\begin{align}
  E
  &=
  \min_{\bar{\boldsymbol{\phi}}}
  \left[
    \boldsymbol\phi^\top
    \begin{pmatrix}
      -\Re\left\{ \mathbf{V} \mathbf{V}^\text{H} \right\} & +\Im\left\{ \mathbf{V} \mathbf{V}^\text{H} \right\} \\
      +\Im\left\{ \mathbf{V} \mathbf{V}^\text{H} \right\} & +\Re\left\{ \mathbf{V} \mathbf{V}^\text{H} \right\}
    \end{pmatrix}
    \bar{\boldsymbol\phi}
    \right.
    \nonumber\\&
    \qquad\qquad\quad
    \left.
    +
    \begin{pmatrix}
      - 2 \Re \left\{ \mathbf{h}_d^\top \mathbf{V}^\text{H} \right\}
      &
      + 2 \Im \left\{ \mathbf{h}_d^\top \mathbf{V}^\text{H} \right\}
  \end{pmatrix}
    \bar{\boldsymbol\phi}
  \right]
  \nonumber\\&
  \triangleq
  \min_{\boldsymbol{\phi}}
  \left[
    \bar{\boldsymbol\phi}^\top
    \mathbf{J}_\text{q}
    \bar{\boldsymbol\phi}
  +
  \boldsymbol{\lambda}^\top_\text{q}
  \bar{\boldsymbol\phi}
  \right]
  ,
  \\
  &\text{s.t.} \quad
  \bar{\boldsymbol\phi} \in
  \left\{ +1, -1 \right\}^{2N_\text{RIS} \times 1}
  ,
  \label{eq:E_su_binary}
\end{align}
\end{subequations}
where
the mutual interaction matrix and
the external magnetic field vector
for quaternary phase shift on RIS,
denoted by~$\mathbf{J}_\text{q} \in \mathbb{R}^{2N_\text{RIS} \times 2N_\text{RIS}}$ and~$\boldsymbol{\lambda}_\text{q} \in \mathbb{R}^{2N_\text{RIS} \times1}$, respectively,
are expressed as
\begin{align}
\mathbf{J}_\text{q}
&=
\begin{pmatrix}
-\Re\left\{
\begin{aligned}
&\operatorname{diag}(\mathbf{f})\,\mathbf{G}
\\[-1mm]
&\times \mathbf{G}^\mathrm{H}\operatorname{diag}(\mathbf{f})^\mathrm{H}
\end{aligned}
\right\}
&
+\Im\left\{
\begin{aligned}
&\operatorname{diag}(\mathbf{f})\,\mathbf{G}
\\[-1mm]
&\times \mathbf{G}^\mathrm{H}\operatorname{diag}(\mathbf{f})^\mathrm{H}
\end{aligned}
\right\}
\\[3mm]
+\Im\left\{
\begin{aligned}
&\operatorname{diag}(\mathbf{f})\,\mathbf{G}
\\[-1mm]
&\times \mathbf{G}^\mathrm{H}\operatorname{diag}(\mathbf{f})^\mathrm{H}
\end{aligned}
\right\}
&
+\Re\left\{
\begin{aligned}
&\operatorname{diag}(\mathbf{f})\,\mathbf{G}
\\[-1mm]
&\times \mathbf{G}^\mathrm{H}\operatorname{diag}(\mathbf{f})^\mathrm{H}
\end{aligned}
\right\}
\end{pmatrix},
\label{eq:J_q}
  \\
  \boldsymbol{\lambda}_\text{q}^\top
  &=
  \begin{pmatrix}
    -2 \Re \left\{ \mathbf{h}_d^\top \mathbf{G}^\text{H} \text{diag} \left( \mathbf{f} \right)^\text{H} \right\}
    &
    +2 \Im \left\{ \mathbf{h}_d^\top \mathbf{G}^\text{H} \text{diag} \left( \mathbf{f} \right)^\text{H} \right\}
  \end{pmatrix}
  .
  \label{eq:lambda_q}
\end{align}

\section{Spin-Size Reduction For Large-Scale Ising Optimization}
\label{sec:size_reduction}

In the previous section,
we formulated the discrete phase shift optimization problem for RISs as an Ising model suitable for CIM.
In this formulation,
the number of required spins increases with the number of RIS elements.
For binary phase shift, the number of spins equals the number of RIS elements~$N_\text{RIS}$,
whereas that becomes $2N_\text{RIS}$ for quaternary phase shift so as to assign the I- and Q-components of the reflection coefficient to separate spins.
Current hardware CIM implementation supports about $50,000$ spins~\cite{honjo2021100},
and RISs with larger aperture or higher carrier frequencies are expected to exceed the limit of available spins,
as they require a larger number of reflecting elements.
Moreover,
in practical CIM implementations,
multiple combinatorial optimization problems can be embedded within a single pulse train composed of $10^4$~pulses,
and reducing the problem size allows more problems to be embedded simultaneously.
For these reasons, reducing the problem size prior to CIM execution is highly beneficial,
even if the obtained result slightly deviates from the optimal solution.

In this section, we propose a spin-size reduction method for CIM specifically tailored for discrete phase optimization of RISs over LoS channels
by analyzing the structure of the Ising Hamiltonian.

\subsection{Dominance of the Magnetic Field Term}

Recall that the optimization problem is expressed by~\eqref{eq:ising_hamiltonian},
the external magnetic field~$\lambda_i$ is computed based on the direct channel between BS and UE~$\mathbf{h}_d$,
and it becomes zero in NLoS environment
from~\eqref{eq:lambda_b} and~\eqref{eq:lambda_q}.
In free-space propagation,
the direct BS-UE path experiences a single path-loss,
whereas the reflected path via the RIS experiences two multiplicative path-loss components (BS-RIS and RIS-UE) and typically longer geometric distances,
i.e., $d_{\mathrm{BS\text{-}UE}} < d_{\mathrm{BS\text{-}RIS}} + d_{\mathrm{RIS\text{-}UE}}$.
As a result, the external magnetic field term $\lambda_i$ is generally much larger than
the mutual interaction term~$J_{ij}$ as
\begin{align}
|\lambda_i| \gg  |J_{ij}|.
\label{eq:bias_dominance}
\end{align}
This relation is especially pronounced under LoS conditions.
In such cases,
the contribution of the mutual interaction terms to the final spin configuration becomes negligible,
and the $i$th external magnetic field~$\lambda_i$ alone dictates the optimal value of the $i$th spin~$\sigma_i$.

If we ignore the interaction terms for a moment,
the optimal spin that minimizes~\eqref{eq:ising_hamiltonian} is obtained in closed form as
\begin{align}
\sigma_i^\star =
\begin{cases}
+1, & \lambda_i < 0, \\
-1, & \lambda_i \ge 0.
\end{cases}
\label{eq:sigma_closed_form}
\end{align}
This means that many spins have an almost predetermined optimal value even before running the CIM.
Surely, the effects of $J_{ij}$ can flip the predetermined spins.
However, if the energy scale of $\lambda_i$ is larger than that of the $J_{ij}$ term,
where the energy of the mutual term should take into account the summation such as $\sum_j \left| J_{ij} \right|$,
some of spins remain in the same direction.
Such spins do not need to be explicitly included in the CIM, and removing them directly reduces the dimensionality of the optimization problem.

Let $\mathcal{S}$ denote the set of indices corresponding to RIS elements
that can be removed in the CIM optimization due to~\eqref{eq:bias_dominance},
and the optimal state of the $i$th spin for~$i \in \mathcal{S}$ is effectively determined by~\eqref{eq:sigma_closed_form}.
After identifying all such spins, we reconstruct a reduced Ising model by:
\begin{enumerate}
\item Remove the rows and columns of the matrix $\mathbf{J}$ corresponding to the indices in $\mathcal{S}$.
\item Insert the predetermined values $\sigma_i^\star$ into the objective as constant terms.
\item Calculate the effective $\boldsymbol{\lambda}$ after removing the spins.
\end{enumerate}

By using predetermined spins~$\sigma^\star_i$ and their indices set~$\mathcal{S}$,
the Ising Hamiltonian in~\eqref{eq:ising_hamiltonian} can be rewritten as
\begin{align}
H ({\boldsymbol{\sigma}})
&=
\sum_{i\in\mathcal{S}^c}\sum_{j\in\mathcal{S}^c} J_{ij}\sigma_i\sigma_j
+\sum_{i\in\mathcal{S}}\sum_{j\in\mathcal{S}} J_{ij}\sigma^\star_i\sigma^\star_j
\nonumber\\&\quad
+
\sum_{i\in\mathcal{S}}\sum_{j\in\mathcal{S}^c} J_{ij}\sigma^\star_i\sigma_j
+
\sum_{i\in\mathcal{S}^c}\sum_{j\in\mathcal{S}} J_{ij}\sigma_i\sigma^\star_j
\nonumber\\&\quad
+
\sum_{i\in\mathcal{S}^c}\lambda_i\sigma_i
+
\sum_{i\in\mathcal{S}}\lambda_i\sigma^\star_i
,
\nonumber\\
&=
\sum_{i\in\mathcal{S}^c}\sum_{j\in\mathcal{S}^c} J_{ij}\sigma_i\sigma_j
\nonumber\\&\quad
+
\sum_{i\in\mathcal{S}^c} \left( \lambda_i + 2 \sum_{j\in\mathcal{S}} J_{ij} \sigma^\star_j \right) \sigma_i
+ C
,
\label{eq:reduced_hamiltonian}
\end{align}
where $\mathcal{S}^c$ denotes the complement of~$\mathcal{S}$,
and $C$ absorbs the constant contributions from the removed spins.

Let $\tilde{\boldsymbol\sigma} \triangleq \boldsymbol\sigma_{\mathcal{S}^c}$ denote the subvector of~$\boldsymbol\sigma$ indexed by~$\mathcal{S}^c$.
The resulting Ising Hamiltonian with spin size reduction is expressed as
\begin{align}
  {H}_{\mathrm{reduced}}
  \left( \boldsymbol\sigma_{\mathcal{S}^c} \right)
  &=
  \sum_{i \in \mathcal{S}^c} \sum_{j \in \mathcal{S}^c} J_{ij} \sigma_i \sigma_j
  +
  \sum_{i \in \mathcal{S}^c} \tilde{\lambda}_i \sigma_i
  \nonumber\\&
  =
  \tilde{\boldsymbol\sigma}^\top \tilde{\mathbf{J}} \tilde{\boldsymbol\sigma}
  +
  \tilde{\boldsymbol\lambda}^\top \tilde{\boldsymbol\sigma}
  ,
  \label{eq:reduced_hamiltonian_2}
\end{align}
where $\tilde{\mathbf{J}} \triangleq \mathbf{J}_{\mathcal{S}^c, \mathcal{S}^c}$ is a submatrix of $\mathbf{J}$ obtained by restricting both rows and columns to indices in~$\mathcal{S}^c$,
and $\tilde{\boldsymbol\lambda}$ is a vector whose $i$th component $\tilde{\lambda}_i$ is given by~$\tilde{\lambda}_i = \lambda_i + 2 \sum_{j\in\mathcal{S}} J_{ij} \sigma^\star_j$.

Although this size reduction approach relies on the dominance of the external magnetic term,
we emphasize that this assumption naturally holds for RISs deployed in LoS environments
since the reflected signal through the RIS experiences two propagation losses,
i.e.,~$\left| h_{d,k} \right|^2 \gg \left| f_n \phi_n G_{n,k} \right|^2$ as given in~\eqref{eq:hd_element} and~\eqref{eq:cascaded_channel}.
Therefore,
this technique provides a practical and effective mechanism for scalable CIM-based optimization for future RIS deployments.

\subsection{Criterion for Spin Removal}

The challenge is determining which spins are safely removable without significantly impacting solution quality.
In this work,
we employ a threshold-based spin classification method,
as described below.

Assuming that the initial value of each spin is determined by the external magnetic term~$\lambda_i$
according to~\eqref{eq:sigma_closed_form},
the $i$th spin may flip when the mutual interaction~$J_{ij}$ is taken into account
if the following condition is satisfied.
\begin{align}
  T_i \triangleq -\text{sgn} \left( \lambda_i \right) \left( \lambda_i + 2 \sum_{j=1, j\neq i}^{N} J_{ij} \text{sgn} \left( \lambda_j \right) \right)
  > 0
  ,
  \label{eq:flip_condition}
\end{align}
where $\text{sgn} \left( \cdot \right)$ represents the sign function that returns the sign of a number.
In practice,
the final spin configuration is determined by the interactions among spins,
and thus the flipping condition cannot be evaluated independently for each spin as in~\eqref{eq:flip_condition}.
In this work,
as a reference criterion,
we adopt the maximum value of~$T_i$ over all spin indices as a reference threshold given by
\begin{align}
T_{\max} = \max_i T_i.
\end{align}
The $i$th spin is considered as \emph{predetermined} if $\left| T_i \right| > T_{\max}$,
i.e.,~$i \in \mathcal{S}$,
and its value is then fixed deterministically based on~\eqref{eq:sigma_closed_form}.
In this case,
the set of predetermined indices corresponding to RIS elements is expressed as
\begin{align}
  \mathcal{S} := \left\{ i \in \{1,\dots,n\} \middle| \left| T_i \right| > T_\text{max} \right\}
  .
  \label{eq:set_S}
\end{align}

\section{Simulation Results}

The performance of discrete phase shifts by using the proposed CIM-based approach is evaluated through numerical simulations in this section.

\begin{figure}[tb]
  \centering
  \includegraphics[width = \hsize, clip]{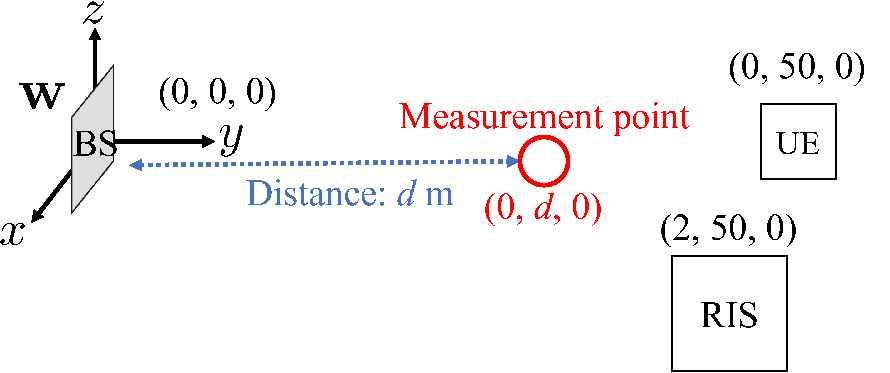}
  \caption{
    The geometric evaluation model of RIS-assisted single user downlink communication.
  }
  \label{system_geo}
\end{figure}

The geometric evaluation model is illustrated in Fig.~\ref{system_geo},
and the simulation parameters are listed in Table~\ref{table:simulation_parameters}.
In a three-dimensional coordinate system,
a BS is located at the origin $(0,0,0)$,
and a UE is placed $50$~m away along the $y$-axis direction at coordinates~$(0,50,0)$.
The BS is equipped with a square-shaped uniform panel array oriented towards the positive $y$-axis direction,
directly facing the UE.
The square-shaped RIS is located near the UE at coordinates~$(2,50,0)$,
with its front side oriented toward the positive $x$-axis direction,
thus facing the UE.
Under this environment,
each design method aims to optimize the discrete phase for a UE positioned at coordinates~$(0,50,0)$.

In the subsequent simulations,
we evaluate the system performance in terms of the channel gain,
which is the optimization metric in the objective function of~\eqref{eq:E_2},
defined as
\begin{align}
  \text{channel gain [dB]}
  &=
  10 \log_{10} \left(
  \left\| \mathbf{h}^\top \right\|^2
  \right)
  .
  \label{eq:channel_gain}
\end{align}
This channel gain
is measured at positions defined by coordinates~$(0,d,0)$, where $d$ represents the distance along the $y$-axis.

\begin{table}[tb]
  \begin{center}
    \caption{Simulation Parameters.}
    \footnotesize
  \begin{tabular}{c|c}
    Parameter & Value \\ \hline \hline
    Carrier frequency & 28.0~GHz \\
    Wavelength~$\lambda$ & 10.71~mm \\
    Number of RIS phase shift levels~$L$ & 2 (binary), 4 (quaternary) \\
    Side length of RIS & 0.4~m, 0.6~m, 0.8~m \\
    Number of RIS elements~$N_\text{RIS}$ & 5476, 12254, 22201 \\
    Number of BS antennas~$N_\text{BS}$ & 64 \\
    Number of UE antennas & 1 \\
    Element spacing of RIS elements & $\lambda /2$ \\
    Element spacing of BS antennas & $\lambda /2$ \\
    Coordinates of BS & (0, 0, 0) \\
    Coordinates of UE & (0, 50, 0) \\
    Coordinates of RIS & (2, 50, 0) \\
  \end{tabular}
  \label{table:simulation_parameters}
  \end{center}
\end{table}

\subsection{Hardware Implementation of Coherent Ising Machine}

\begin{figure}[tb]
  \centering
  \includegraphics[width=1\linewidth]{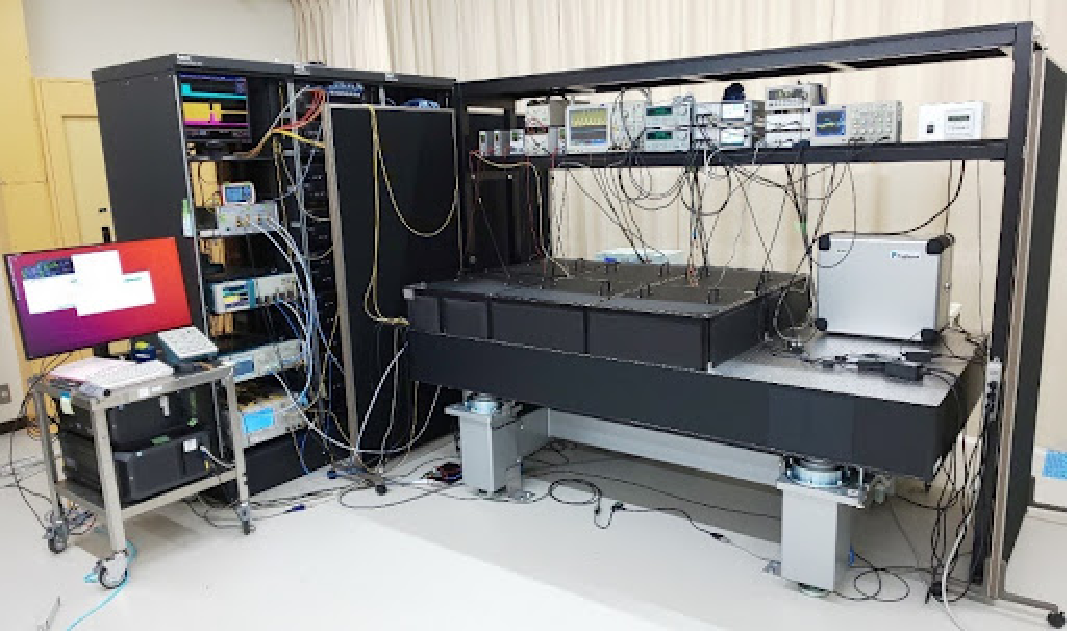}
  \caption{Photograph of the hardware CIM used in this work.}
  \label{cim_photo}
\end{figure}

To evaluate the practical effectiveness of the proposed CIM-based method,
we performed discrete phase shift optimization using a CIM developed by NTT~\cite{honjo2021100}.
Fig.~\ref{cim_photo} shows a photograph of the CIM used in our experiments.
The CIM considered in this study is a physical Ising solver based on time-multiplexed degenerate optical parametric oscillator~(DOPO) pulses with a measurement-feedback architecture.
The detailed hardware configuration and operating conditions of the CIM have been reported in~\cite{takesue2025finding}.
In particular,
the CIM employs an auxiliary-spin scheme to implement effective external fields,
which enables stable optimization of large-scale and dense graph problems.
The CIM can realize fully connected Ising models with up to approximately $50,000$ spins,
where the interaction coefficients are controlled with $8$-bit resolution.
Each annealing process is executed with a fixed duration of approximately $22$~ms,
and the final spin states are obtained from the sign of the DOPO pulse amplitudes.
Unlike software-based emulators such as simulated annealing,
it physically realizes the Ising dynamics through optical and electronic components,
enabling large-scale combinatorial optimization via real hardware operation.
In this section,
all optimization results were obtained
by executing the proposed algorithm directly on this hardware CIM,
thereby demonstrating the feasibility and scalability of the proposed approach
under realistic computational constraints.


\subsection{Comparison with Conventional Approaches}

We first compare the performance achieved by our proposed CIM-based approach
with the existing approaches.
For comparison,
we introduce two conventional approaches,
successive refinement algorithm~\cite{wu2019beamforming} and Fresnel zone-based design~\cite{kitayama2021transparent}.
The successive refinement algorithm is an iterative method
that optimizes the discrete phase shift of each RIS element one by one while keeping the others fixed.
This algorithm features low computational complexity,
scales well with large numbers of reflecting elements,
and achieves near-optimal performance.
Fresnel zone-based design is a technique
whereby each element of the RIS inverts its phase according to
whether it lies within an even-order or an odd-order Fresnel zone.

In the following of this subsection,
we compare the performance for the cases with and without a LoS path
between the BS and the UE,
which is given by~$\mathbf{h}_d^\top$ in~\eqref{eq:h}.

\subsubsection{Non-Line-of-Sight~(NLoS) Channel}

First, we consider the NLoS scenario in which no direct path exists between the BS and UE,
i.e., $\mathbf{h}_d = \mathbf{0}^{N_\text{BS} \times1}$
with $\mathbf{0}$ representing a column zero vector,
meaning signals are received solely through the RIS-assisted path.




\begin{figure}[tb]
  \centering
  \vspace{-3.52mm}
  \subfigure[Overall performance ($d=$0–100 m)]{\includegraphics[width=\hsize]{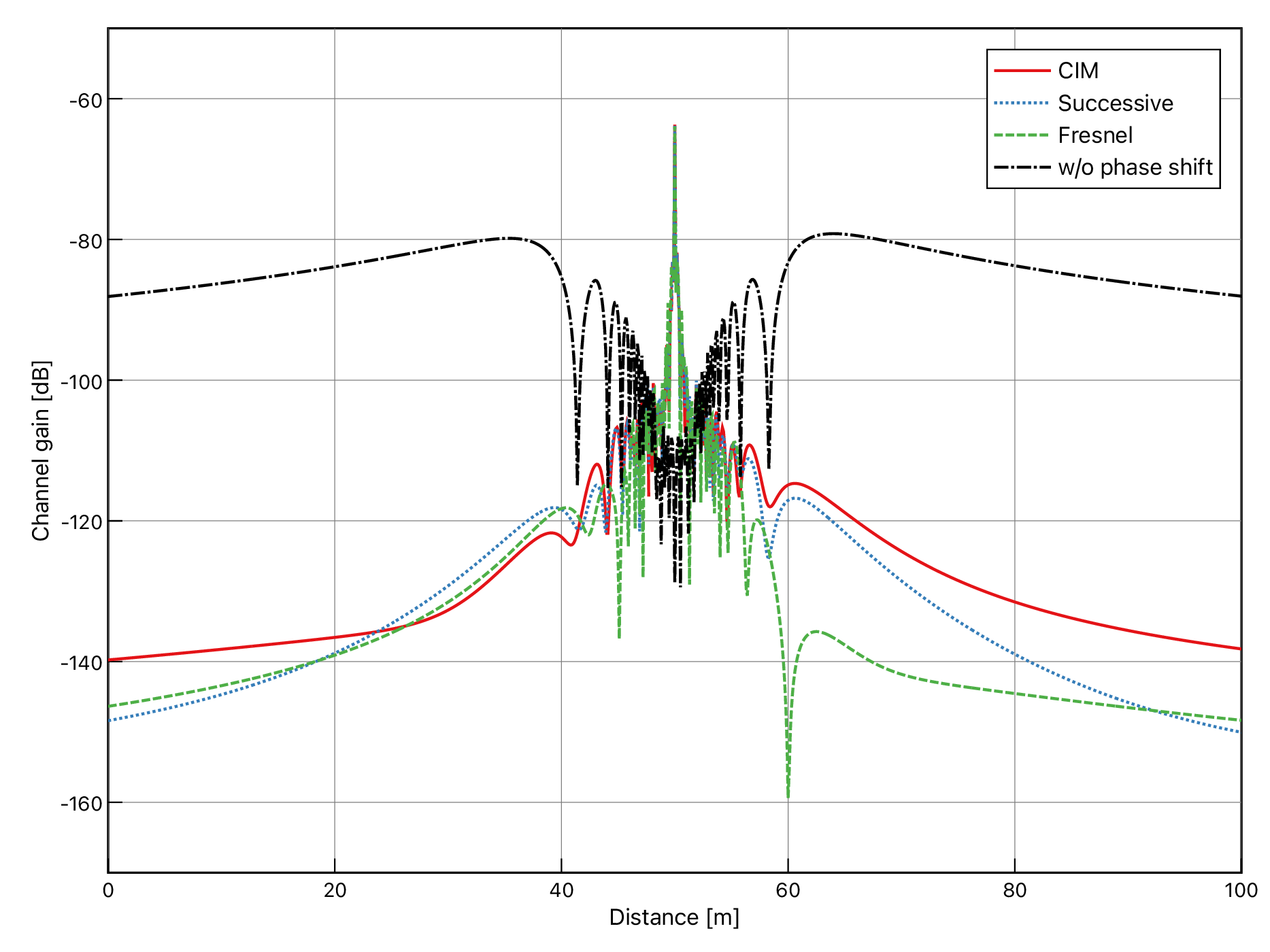}
    \label{su_binary_nlos}}
  \subfigure[Enlarged view around $d=$50 m]{\includegraphics[width=\hsize]{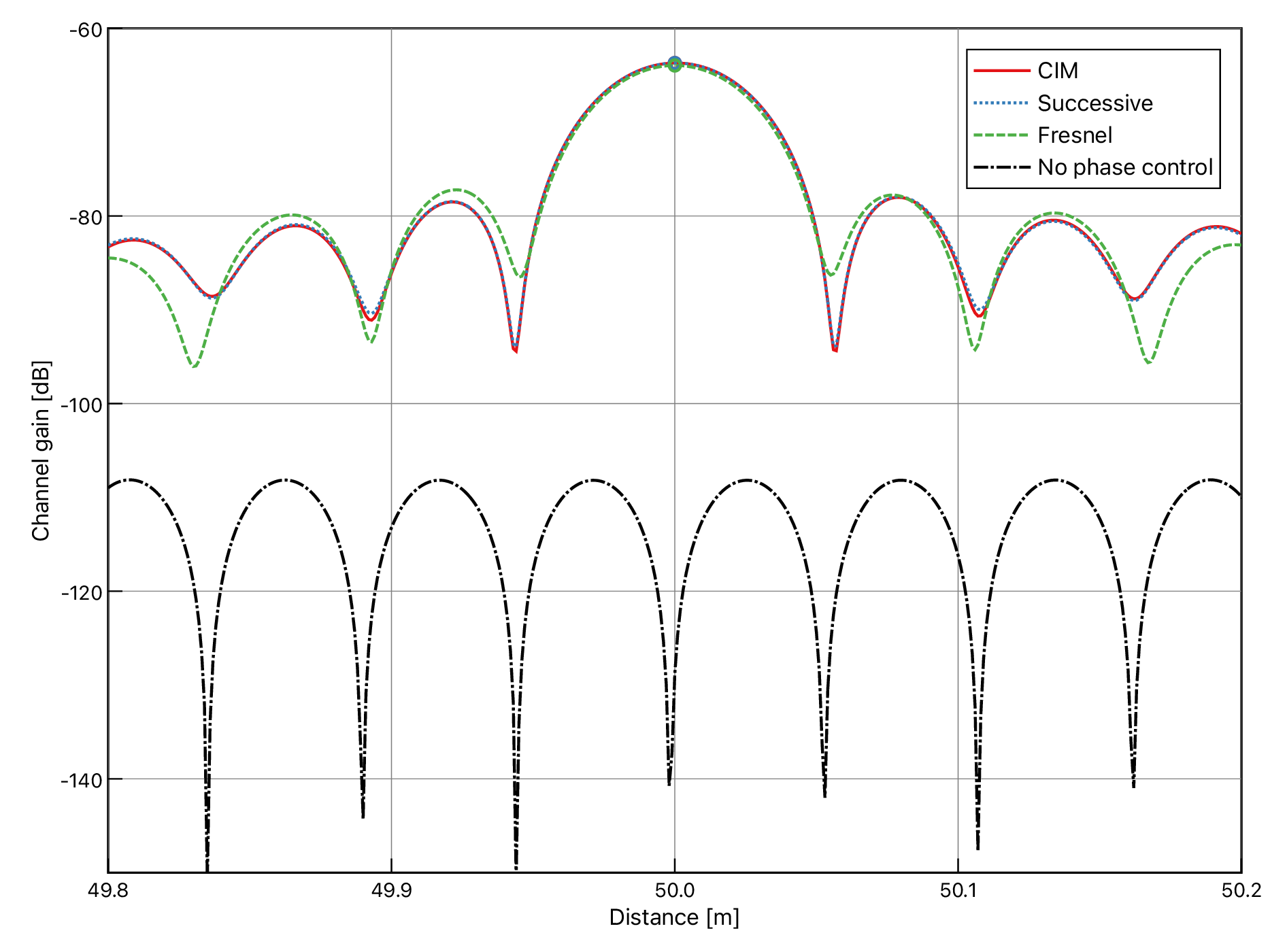}
    \label{new_nlos_comparison_1bit}}
  \caption{
    The channel gain comparison using different binary phase shift schemes of RIS
    over NLoS channel ($N_\text{RIS} = 5476$).
    For reference, the values at~$d = 50$~m are
    $-63.70$~dB, $-63.70$~dB, and $-63.95$~dB for CIM, successive, and Fresnel, respectively.
  }
  \label{nlos_comparison}
\end{figure}

Fig.~\ref{nlos_comparison} shows the results for the case where the RIS comprises $N_\text{RIS} = 5476$ elements (corresponding to a square with a side length of $0.4$~m),
and we consider binary phase shift ($L = 2$).
Hence, the number of possible phase combinations for the RIS elements is $2^{5476}$.
Note that Fig.~\ref{su_binary_nlos} plots the overall performance over the distance range from $0$~m to $100$~m,
while Fig.~\ref{new_nlos_comparison_1bit} shows an enlarged view around the designed distance of $50$~m.
In addition to the conventional successive refinement algorithm and Fresnel zone-based design,
results obtained without phase control,
where all RIS elements fixed at phase 0, i.e., acting merely as a passive reflector,
are included in Fig.~\ref{nlos_comparison}.
Comparing the three discrete phase shift methods at the UE position $(0,50,0)$, both the proposed CIM-based method and successive refinement yield the same received power values of \(-63.70\,\text{dB}\), slightly outperforming the Fresnel-zone-based approach, which achieves \(-63.95\,\text{dB}\).
In addition,
the achievable channel gain by ideal continuous phase shift is $-59.85$~dB.
The gap between CIM and ideal continuous phase shift is $3.85$~dB,
and the performance by CIM and successive refinement is approaching the power loss by binary phase shift of~$3.9$~dB.
Therefore,
the discrete phase shift based on the proposed CIM
works appropriately and effectively.

\subsubsection{Line-of-Sight~(LoS) Channel}

Next, we consider the LoS scenario,
where a direct path between the BS and UE exists.
In this scenario,
the signals reflected by the RIS should be constructively combined in phase and aligned with the direct path from the BS.




\begin{figure}[tb]
  \centering
  \vspace{-3.52mm}
  \subfigure[Overall performance ($d=$0–100 m)]{\includegraphics[width=\hsize]{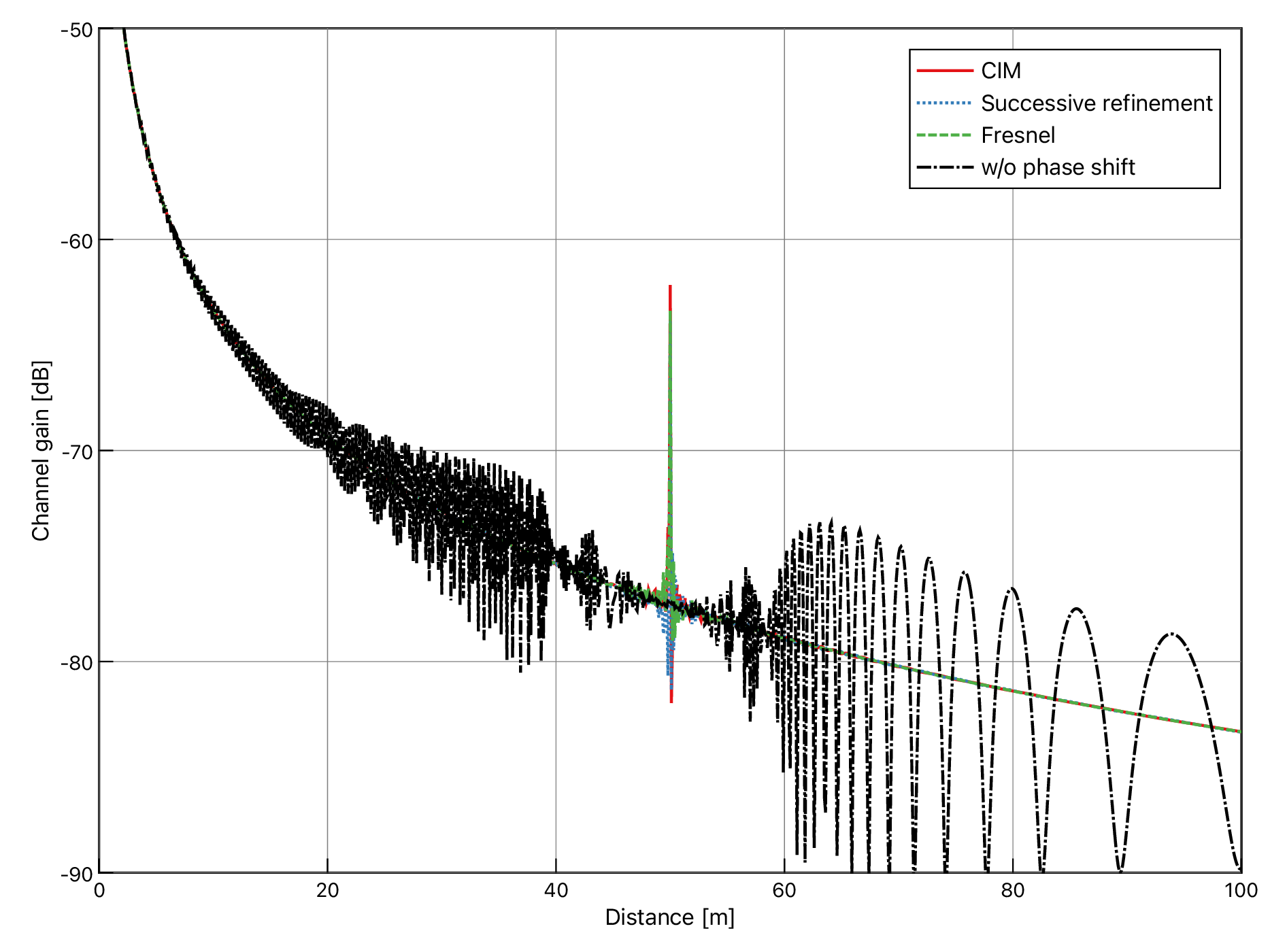}
    \label{su_binary_los}}
  \subfigure[Enlarged view around $d=$50 m]{\includegraphics[width=\hsize]{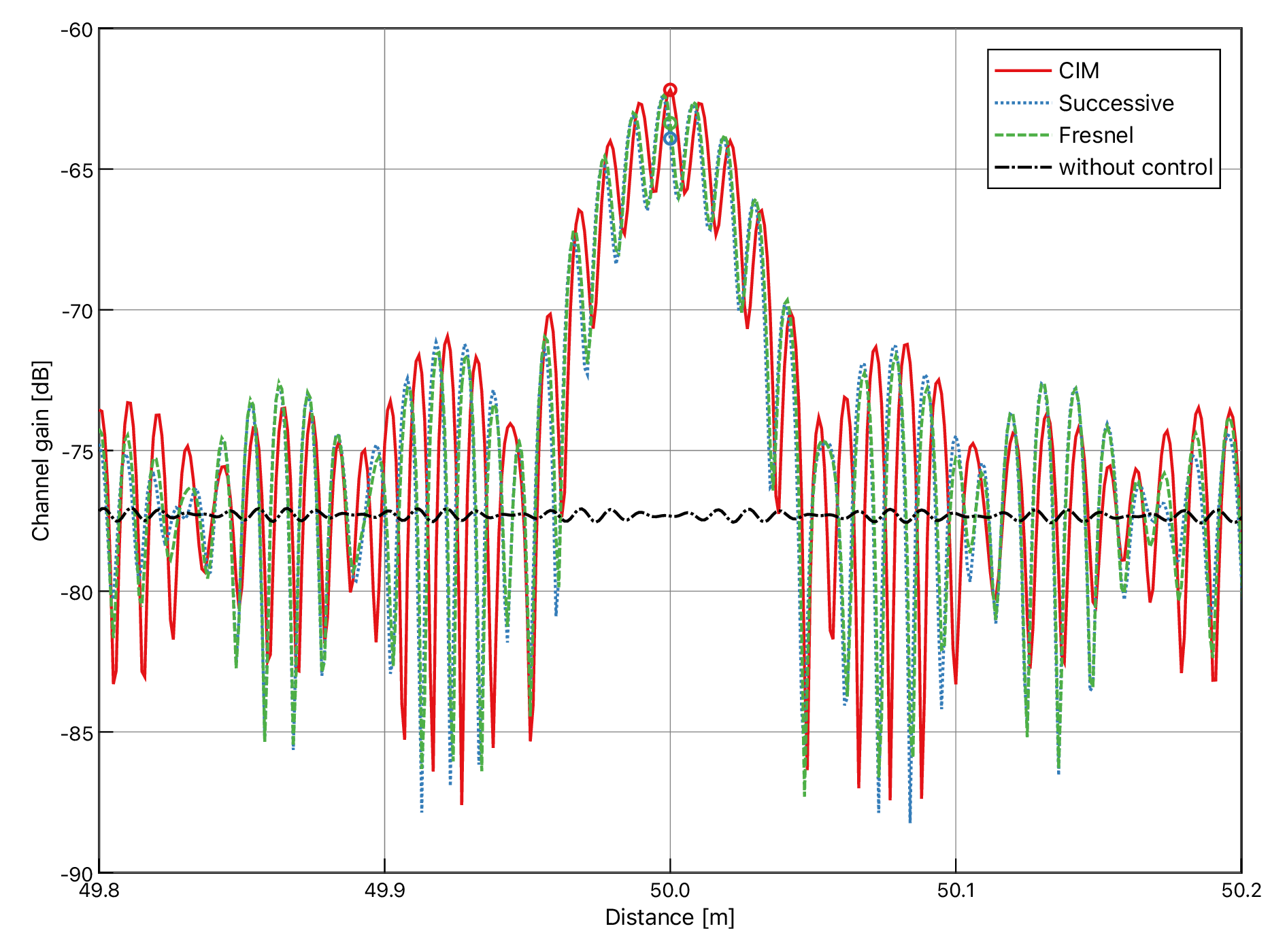}
    \label{new_los_comparison_1bit}}
  \caption{
    The channel gain comparison using different discrete phase shift schemes of RIS
    over LoS channel ($N_\text{RIS} = 5476$).
    For reference, the values at~$d = 50$~m are
    $-62.16$~dB, $-63.84$~dB, and $-63.37$~dB for CIM, successive, and Fresnel, respectively.
  }
  \label{los_comparison}
\end{figure}

Fig.~\ref{los_comparison} shows the results for~$N_\text{RIS}=5476$ obtained by each phase shift method under the LoS scenario.
Here, the simulation parameters are identical to those in Fig.~\ref{nlos_comparison},
except that the channel assumption is changed from NLoS to LoS.
In addition to the proposed CIM-based method,
results from conventional methods such as successive refinement and the Fresnel-based method are also presented for comparison.
Examining the channel gain at the UE position for each design approach,
the CIM achieves $-62.16$~dB,
successive refinement achieves $-63.37$~dB,
and the Fresnel zone-based design achieves $-63.84$~dB.
Unlike the NLoS scenario, in the LoS case,
the proposed CIM approach slightly but clearly outperforms the other schemes.
The inferior performance of the Fresnel method can be attributed to
the lack of the phase alignment with the direct path,
as it only considers the RIS-assisted path.
Similarly, successive refinement, being a suboptimal iterative algorithm,
may converge to a discrete phase configuration considerably different from the optimal solution.
On the other hand, the formulation of the CIM inherently takes into account the direct path.
Consequently,
the proposed CIM approach successfully obtains a highly effective discrete phase combination on the RIS.

The above result in Fig.~\ref{los_comparison} addresses a problem attributed to the external magnetic field term~$\boldsymbol\lambda$,
which has required further investigation
in previous studies on CIM~\cite{takesue2025finding}.
Moreover,
the LoS environment represents a scenario in which
the external magnetic field term becomes more dominant than the mutual interaction term.
Such a scenario has not been sufficiently explored in conventional CIM research.
In this study,
we demonstrate that
the CIM operates appropriately and is capable of obtaining reasonable solutions even under such conditions.
These findings confirm the applicability
and facilitate its application to larger-scale RIS systems under constraints on the number of spins.

\subsection{RIS Size Comparison}

In this subsection, we investigate the impact of the RIS size, i.e., the number of reflecting elements, on the achievable channel gain under both NLoS and LoS channel conditions.
In both cases, three RIS size configurations are considered,
and corresponding side lengths of the RIS are $0.4$~m, $0.6$~m, and $0.8$~m.
We here consider the same element spacing of~$\lambda/2$,
and
the numbers of RIS elements depending only on their aperture are given by
$N_\text{RIS} = 5476$, $N_\text{RIS} = 12254$, and $N_\text{RIS} = 22201$, respectively.
Note that we consider binary phase shift in the following.

\begin{figure}[tb]
  \centering
  \includegraphics[width = \hsize, clip]{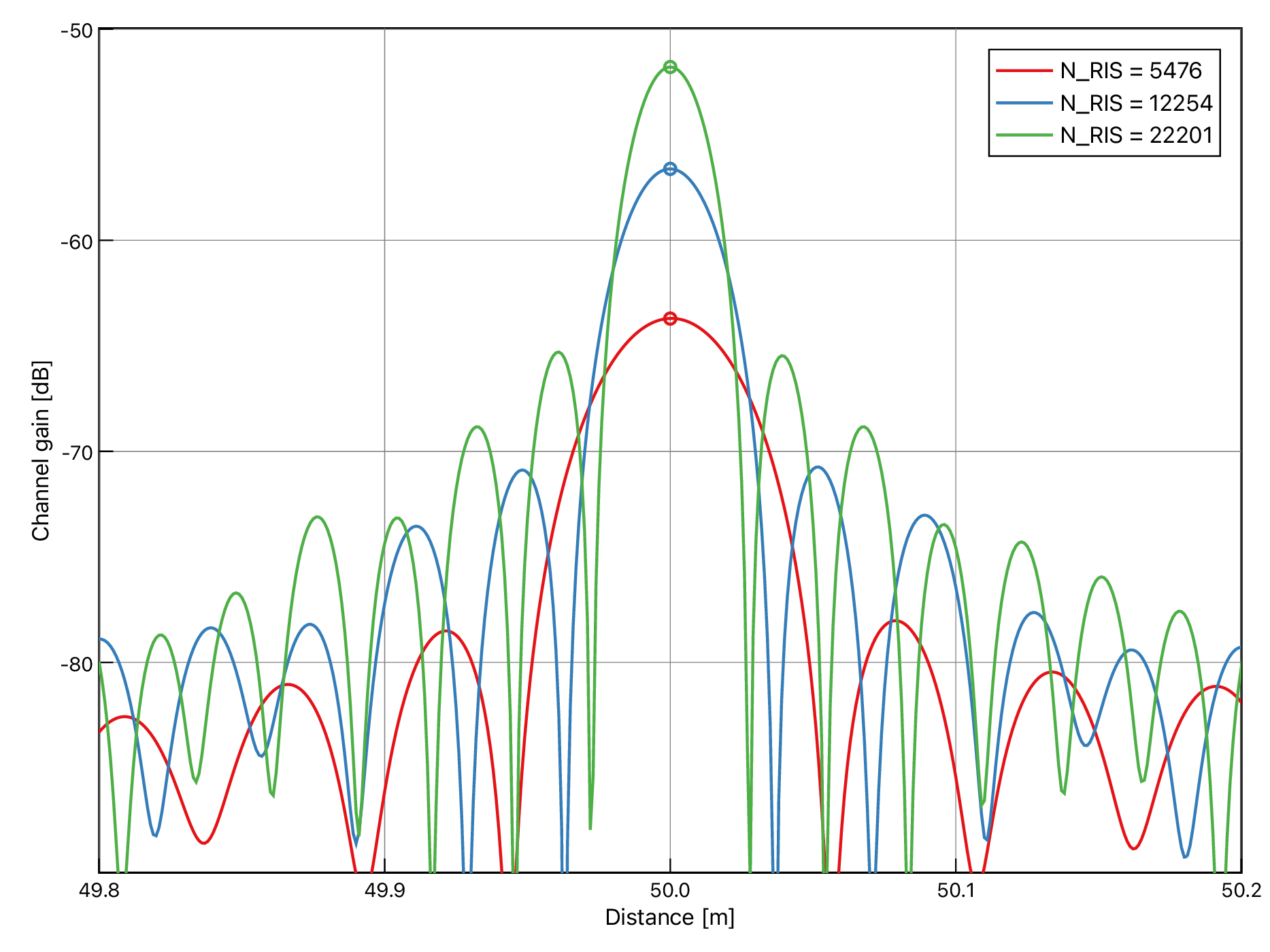}
  \caption{
    The channel gain via RIS with binary discrete phase shifts based on CIM over NLoS channels
    for different RIS size ($N_\text{RIS} = 5476, 12544$, and $22201$).
    For reference, the values at~$d = 50$~m are
    $-63.70$~dB, $-56.62$~dB, and $-51.79$~dB for $N_\text{RIS}=5476$, $N_\text{RIS}=12544$, and $N_\text{RIS}=22201$, respectively.
  }
  \label{nlos_binary_size}
\end{figure}

Fig.~\ref{nlos_binary_size} shows the channel gain of the proposed CIM method under NLoS conditions
for different RIS sizes.
The resulting channel gains at the UE location
for three RIS configurations of~$N_\text{RIS}=5476$, $N_\text{RIS}=12254$, and $N_\text{RIS}=22201$
are $-63.70$~dB, $-56.62$~dB, and $-51.79$~dB, respectively.
Since the ratios of the RIS areas relative to $N_\text{RIS}=5476$ are $1$, $2.29$, and $4.05$ respectively,
theoretical expectation for gain improvements by using larger aperture RISs
are about $3.60$~dB and $6.07$~dB for $N_\text{RIS}=12254$ and $N_\text{RIS}=22201$ compared to the case of $N_\text{RIS}=5476$.
As the received power involves applying this RIS array gain twice,
i.e., once before and once after reflection,
the achievable channel gains for $N_\text{RIS}=12254$ and $N_\text{RIS}=22201$
are given by $7.20$~dB and $12.14$~dB, respectively,
compared to that with~$N_\text{RIS}=5476$.
The above relative gain improvements observed by experiments agree well with the theoretical expectations.
Thus, we can conclude that
the proposed CIM based discrete phase shift method functions appropriately even for very large-scale RIS.

\begin{figure}[tb]
  \centering
  \includegraphics[width = \hsize, clip]{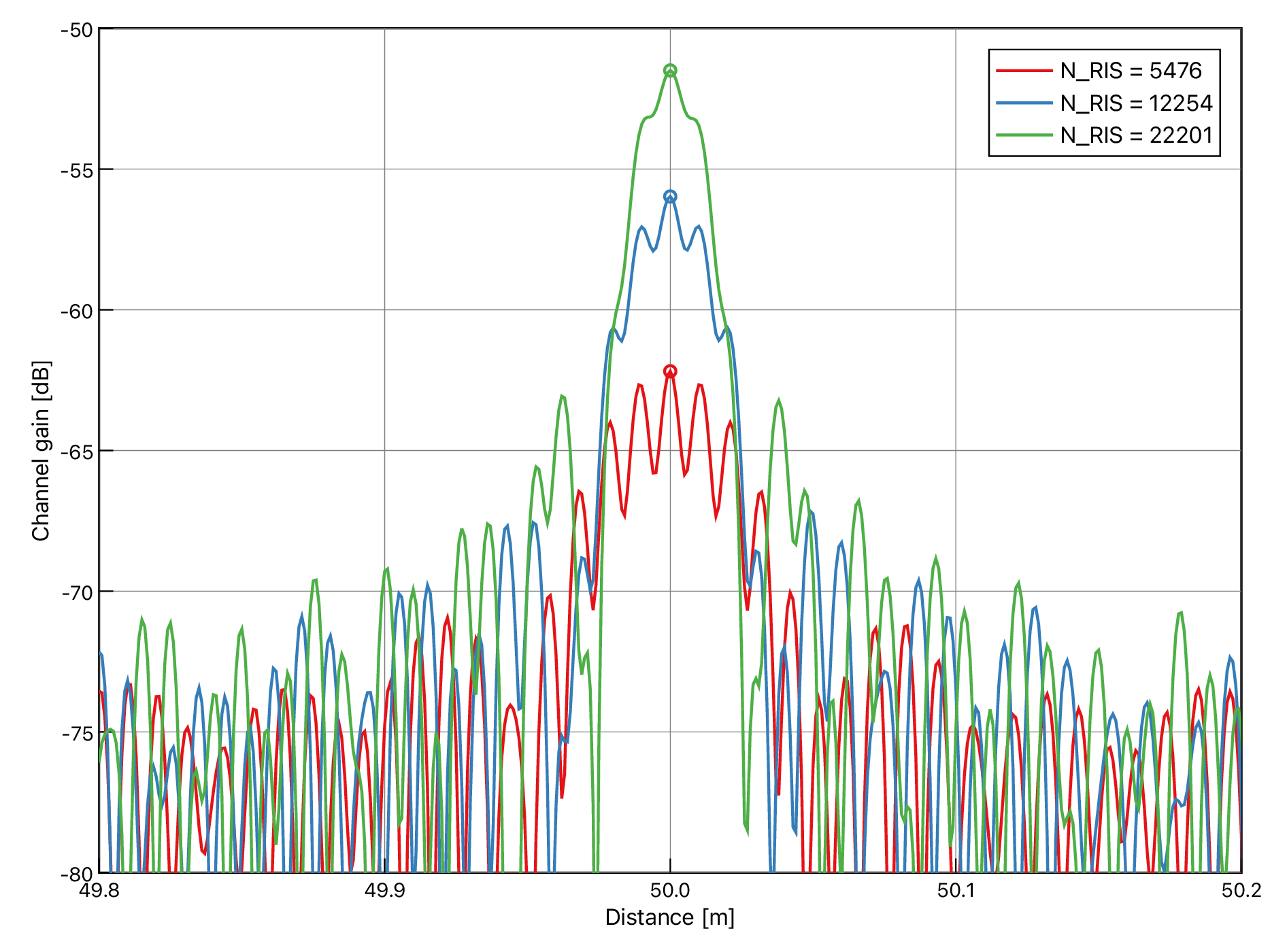}
  \caption{
    The channel gain via RIS with binary discrete phase shifts based on CIM over LoS channels
    for different RIS size ($N_\text{RIS} = 5476, 12544$, and $22201$).
    For reference, the values at~$d = 50$~m are
    $-62.19$~dB, $-55.97$~dB, and $-51.50$~dB for $N_\text{RIS}=5476$, $N_\text{RIS}=12544$, and $N_\text{RIS}=22201$, respectively.
  }
  \label{los_binary_size}
\end{figure}

Fig.~\ref{los_binary_size} shows the simulation results for the LoS channel case,
where all system parameters are identical to those used in Fig.~\ref{nlos_binary_size}.
In Fig.~\ref{los_binary_size},
the channel gains achieved with the RIS configurations $N_\text{RIS} = 5476$, $N_\text{RIS} = 12254$, and $N_\text{RIS} = 22201$ are $-62.19$~dB, $-55.97$~dB, and $-51.50$~dB, respectively.
Although these gains are smaller than the theoretically expected improvements of $7.20$~dB and $12.14$~dB
for $N_\text{RIS} = 12254$ and $N_\text{RIS} = 22201$ relative to $N_\text{RIS} = 5476$,
this discrepancy can be attributed to the presence of the direct path in addition to the reflected path through RIS.
Nevertheless,
as in the NLoS scenario,
the experimental results in Fig.~\ref{los_binary_size} confirmed that
the channel gain increases proportionally to the area of the RIS over the LoS channel.
In the LoS environment,
both the direct path between the BS and the UE
and the reflected path via the RIS interfere coherently at the receiver.
As a result,
more rapid spatial oscillations of the channel gain are observed with respect to the distance~$d$,
corresponding to the constructive and destructive interference between the two propagation components.
However,
as the RIS size increases from $N_\text{RIS}=5476$ to $N_\text{RIS}=22201$,
the influence of the RIS-reflected component becomes increasingly dominant compared to the direct path.
Consequently,
the composite beam pattern becomes smoother,
indicating that the large-scale RIS provides stronger and more coherent reflections
that effectively shape the overall received field distribution.

For large-scale RIS such as~$N_\text{RIS}=22201$, 
there would be~$2^{22201}$ possible phase combinations for exhaustive search,
an astronomically large number beyond the computational capabilities of practical computers.
However,
the present system with parameters~$N_\text{RIS}=22201$,
a RIS aperture of about~$0.8$~m and the carrier frequency of $28$~GHz,
is practically realistic
since the hardware with the similar parameters have been implemented in~\cite{kitayama2021transparent}.
Therefore,
the proposed discrete phase shift based on the CIM
can be regarded as a groundbreaking approach capable of achieving high performance even for large-scale RISs.

\subsection{Comparison Between Binary and Quaternary Phase Shift}

\begin{figure}[tb]
  \centering
  \includegraphics[width = \hsize, clip]{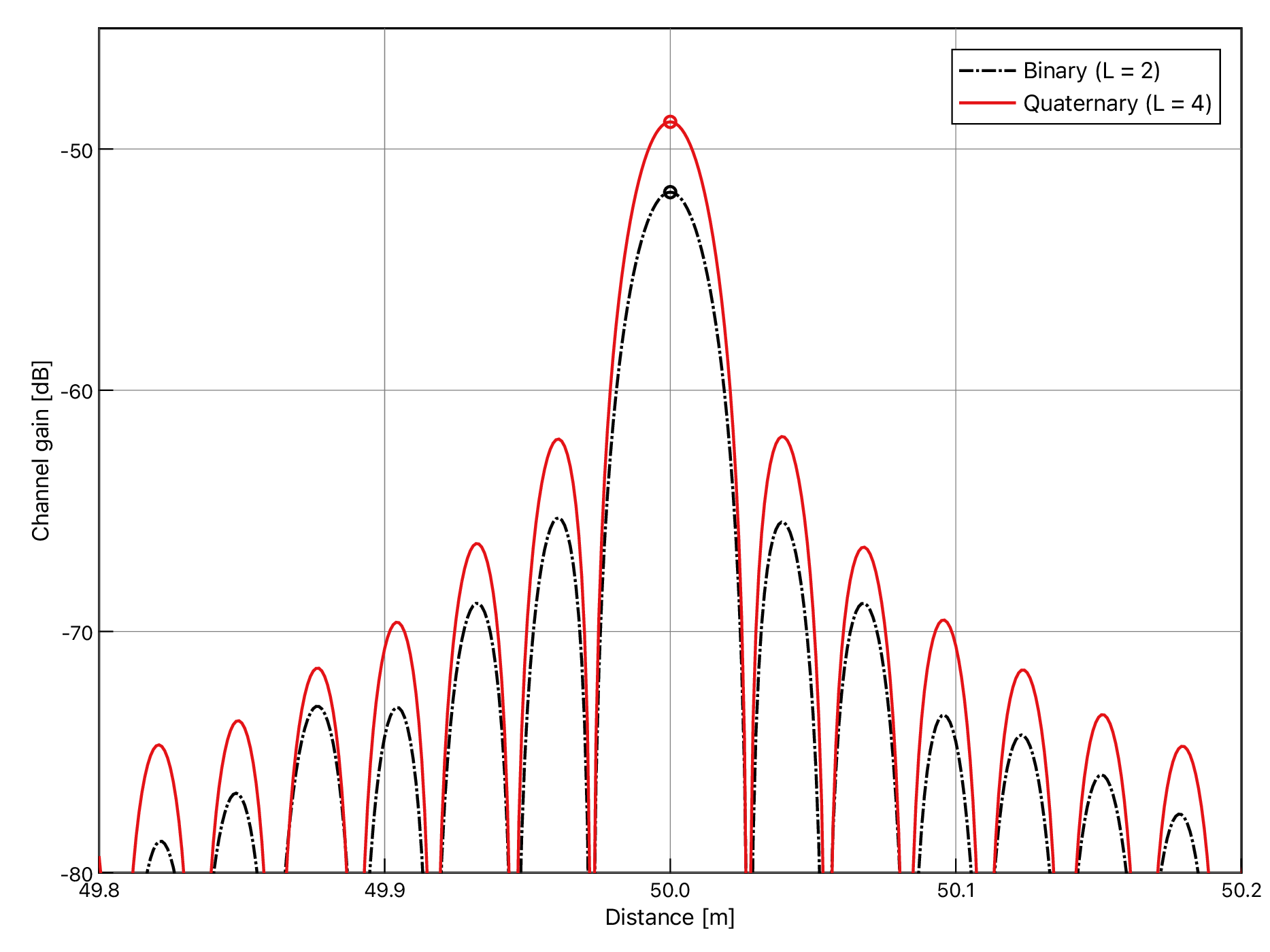}
  \caption{
    The channel gain via RIS with binary and quaternary discrete phase shifts based on CIM over LoS channels ($N_\text{RIS} = 22201$).
    For reference, the values at~$d = 50$~m are
    $-51.79$~dB and $-48.88$~dB for binary and quaternary, respectively.
  }
  \label{new_nlos_1bit_2bit}
\end{figure}

\begin{figure}[tb]
  \centering
  \includegraphics[width = \hsize, clip]{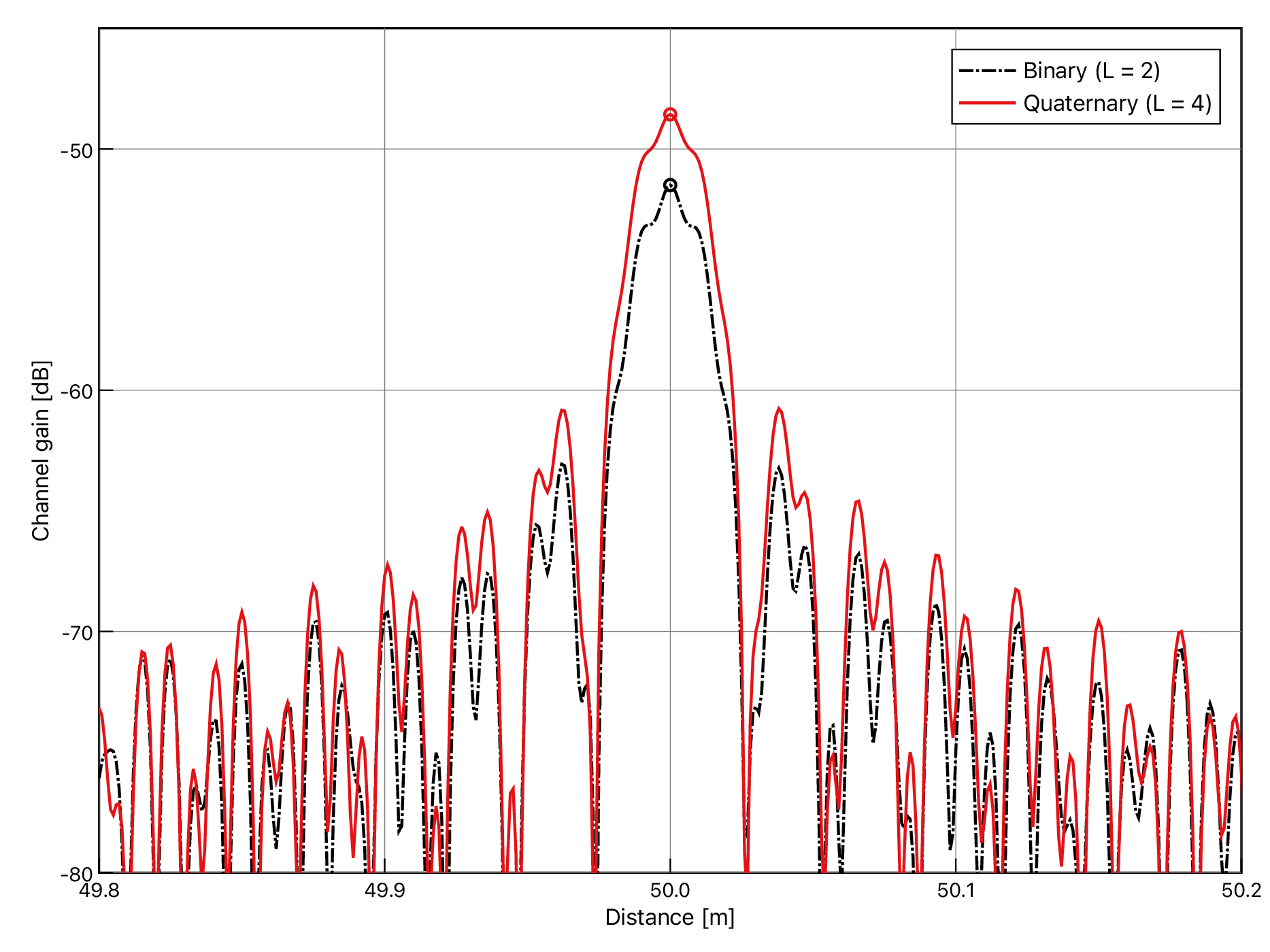}
  \caption{
    The channel gain via RIS with binary and quaternary discrete phase shifts based on CIM over LoS channels
    ($N_\text{RIS} = 22201$).
    For reference, the values at~$d = 50$~m are
    $-51.50$~dB and $-48.57$~dB for binary and quaternary, respectively.
  }
  \label{new_los_1bit_2bit}
\end{figure}

We next consider quaternary phase shift.
Figs.~\ref{new_nlos_1bit_2bit} and~\ref{new_los_1bit_2bit} compare the results between the binary and quaternary phase shift schemes under NLoS and LoS environments, respectively, with~$N_\text{RIS} = 22201$.

In Fig.~\ref{new_nlos_1bit_2bit} for the NLoS case,
the UE receives the signal exclusively through the RIS-reflected path.
As a result,
the spatial structure of the received channel gain directly corresponds to the beam pattern generated by the RIS.
The figure clearly shows that the use of quaternary phase shifts yields approximately a $3$~dB improvement over the binary case across all distances~$d$,
which agrees with
the theoretical prediction for the channel gain with increasing $L$ from $2$ to $4$.
This consistent improvement demonstrates that
the proposed CIM-based optimization effectively determines high-resolution discrete phase configurations
even when the RIS has only limited quantized states.

The results for the LoS scenario,
shown in Fig.~\ref{new_los_1bit_2bit},
exhibit different characteristics.
In this case,
the received field is formed by the coherent superposition of the direct BS-UE path and the RIS-reflected path.
For the LoS,
we found the rapid oscillation with respect to the distance~$d$ as mentioned above.
Despite these oscillations,
the quaternary control consistently outperforms the binary control at all distances,
again providing nearly a 3~dB improvement.
The uniformity of this gain across the entire range of~$d$ indicates that
additional phase resolution translates directly into additional coherent combination,
resulting in the gain improvement.

Overall,
both figures highlight that
quaternary phase shift control offers systematic performance benefits over binary control
in both NLoS and LoS environments.
Furthermore,
the proposed CIM-based optimization successfully handles the increased number of discrete states required for quaternary phase shift.
The only additional computational cost is a doubling of the number of spins compared to the binary formulation,
which remains well within the capacity of current CIM hardware.
These results confirm that
the CIM-based approach is sufficiently scalable and effective for practical RIS deployments requiring higher phase resolution.

\subsection{Spin Size Reduction}
\label{sec:simulation_size_reduction}

\begin{figure}[tb]
  \centering
  \includegraphics[width = \hsize, clip]{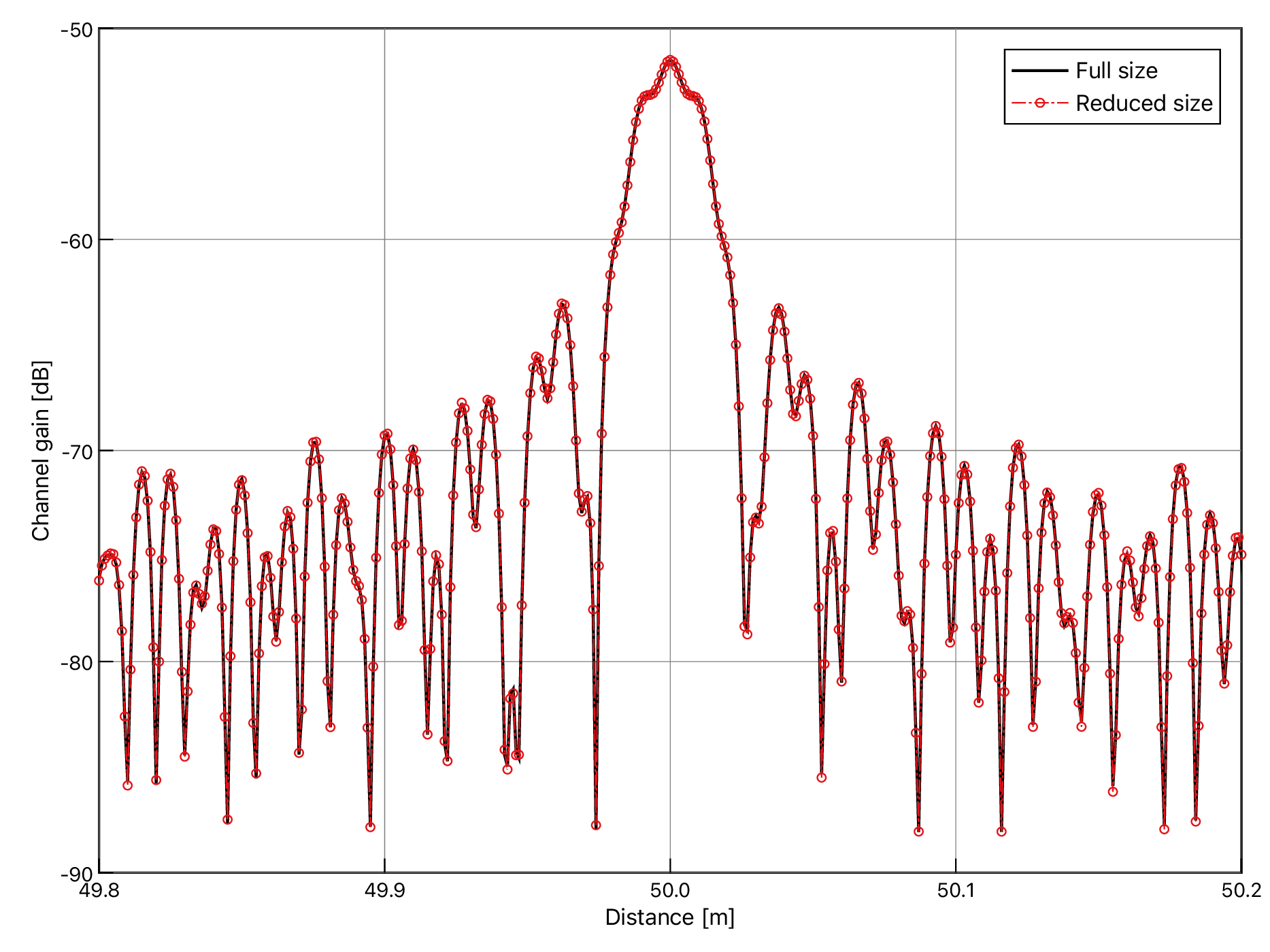}
  \caption{
    Comparison of full-size and reduced-size CIM optimization for binary phase shift control with $N_\text{RIS} = 22201$ under LoS conditions.
    After applying the size reduction,
    the number of spins is reduced to $19277$ while preserving identical channel-gain performance.
    For reference, the values at~$d = 50$~m are
    $-51.50$~dB and $-51.50$~dB for full size and reduced size, respectively.
  }
  \label{ris_bit_reduction}
\end{figure}

Finally,
we investigate the effect of the spin size reduction algorithm in Section~\ref{sec:size_reduction}.
Fig.~\ref{ris_bit_reduction} illustrates the effect of the proposed spin size reduction method for the large RIS with $N_\text{RIS} = 22201$ elements under the LoS environment.
In Fig.~\ref{ris_bit_reduction},
we consider binary phase shift and compare the results obtained by the full Ising model in~\eqref{eq:ising_hamiltonian} with those obtained after applying the reduction algorithm in~\eqref{eq:reduced_hamiltonian_2} in Section~\ref{sec:size_reduction},
which selectively removes spins whose optimal states are determined almost entirely by the external magnetic term only.
As shown in the figure, the channel gain curves obtained from the full model and the reduced model coincide perfectly over the entire range of the distance~$d$.
Both approaches achieve an identical channel gain of $-51.50$~dB at the UE distance~$d = 50$~m,
indicating that our proposed spin size reduction does not degrade the received signal power.

In this particular scenario, the dominance of the direct BS-UE path results in large magnetic fields,
making this algorithm highly effective.
Out of the original $22201$ spins, $19277$ remained after reduction, meaning that $13.2\%$ of the spins were eliminated while preserving the optimized phase configuration.
In this simulation setup,
the RIS is located in close proximity to the UE,
and its contribution to the received signal cannot be neglected.
As a result, the proposed reduction method yields a moderate spin reduction ratio of $13.2\%$.
However,
for users whose received signals are less influenced by the RIS,
a significantly higher reduction ratio can be expected.
In the extreme case where the RIS-assisted path is negligible and only the direct BS-UE channel is present,
a $100\%$ reduction is achieved,
indicating that RIS phase optimization is unnecessary for such UEs.
This spin size reduction demonstrates the practicality of the proposed approach when applied to RIS systems operating in LoS-dominant environments, where the magnetic field term plays a decisive role in determining the optimal spin states.

The observed behavior also provides insights beyond the RIS-specific context.
The reduced model yields equivalent to those of the full model,
indicating that the proposed reduction strategy is applicable to a wide class of Ising-formulated optimization problems.
Whenever the external magnetic term is significantly larger than
the sum of the mutual interaction terms for each spin,
such spin states
are effectively predetermined, and eliminating them does not affect the global optimum.
Thus, the spin size reduction approach offers a general mechanism for drastically shrinking the problem size without sacrificing accuracy.


\section{Conclusion}
\label{sec:conclusion}

This paper proposed a coherent Ising machine~(CIM) based framework for discrete phase shift optimization in large RIS-assisted wireless systems,
and its effectiveness was confirmed by using a real hardware CIM developed by NTT.
By formulating the discrete phase shift problem of RIS as an Ising Hamiltonian,
the CIM was shown to efficiently handle both binary and quaternary phase shift
and to achieve physically consistent beamforming gains for RISs with more than twenty thousand elements.
Experimental simulations demonstrated that the proposed method achieves the theoretical expectation of $3$~dB improvement when moving from binary to quaternary phase shifts and accurately reproduces the expected aperture-scaling behavior in both NLoS and LoS environments.
Furthermore,
a spin size reduction approach was proposed,
exploiting the dominance of external magnetic terms to remove spins
whose optimal states are predetermined in the LoS environment,
which demonstrated about $13.2$~$\%$ spin size reduction for large-scale RIS with 22,201 elements without any performance degradation.
These results confirm that the CIM provides a practical, scalable, and generalizable tool
for discrete phase shift optimization of RIS,
and they indicate strong potential for extending the method to even larger RISs and broader classes of Ising-formulated wireless optimization problems in future work.

\bibliographystyle{IEEEtran}
\bibliography{IEEEabrv,bibhama}
\bstctlcite{BSTcontrol}

\end{document}